\documentclass[%
aip,
 amsmath,amssymb,
 reprint,%
]{revtex4-1}

\usepackage{graphicx}
\usepackage{dcolumn}
\usepackage{bm}
\usepackage{amsthm}
\usepackage{caption}

\usepackage[utf8]{inputenc}
\usepackage[T1]{fontenc}
\usepackage{mathptmx}
\usepackage{etoolbox}

\makeatletter
\def\@email#1#2{%
 \endgroup
 \patchcmd{\titleblock@produce}
  {\frontmatter@RRAPformat}
  {\frontmatter@RRAPformat{\produce@RRAP{*#1\href{mailto:#2}{#2}}}\frontmatter@RRAPformat}
  {}{}
}%
\makeatother

\usepackage{natbib}
\bibpunct[, ]{(}{)}{;}{a}{}{,}

\usepackage{enumerate}
\usepackage{url}
\usepackage{framed}
\usepackage{multirow}
\usepackage{tabularx}
\usepackage{amsfonts}
\usepackage{xspace}
\usepackage{arydshln}
\usepackage[shortlabels]{enumitem}
\usepackage{lipsum}
\usepackage{setspace}
\usepackage{epsf}
\usepackage{psfrag}
\usepackage[modulo]{lineno}
\usepackage{mathtools}
\usepackage{tabulary}
\usepackage{csquotes}
\usepackage{booktabs}
\usepackage{tabularray}
\usepackage{makecell}
\usepackage{MnSymbol}
\usepackage{array}
\usepackage{ulem}
\usepackage{anyfontsize}
\usepackage{hyperref}
\setlist[enumerate]{itemsep=3.5mm}
\usepackage{tablefootnote}
\usepackage{float}
\usepackage{adjustbox}
\usepackage{subfig}

\newcommand{\ex}{{\mathbb E}}

\DeclareMathOperator*{\argminC}{\arg\min}
\usepackage[ruled,linesnumbered]{algorithm2e}
\makeatletter
\renewcommand{\fnum@algocf}{Procedure} 
\makeatother

\theoremstyle{plain}
\newtheorem{theorem}{Theorem}[section]

\theoremstyle{definition}
\newtheorem{definition}[theorem]{Definition}

\theoremstyle{remark}

\begin{document}

\preprint{AIP/123-QED}

\title[Nonlinear Causality]{Nonlinear Causality in Time Series Networks: \\With Application to Motor Imagery vs Execution}
\author{Sipan Aslan}
 \altaffiliation[Also at ]{Department of Econometrics, Van Yuzuncu Yil University.}

\author{Hernando Ombao}%
 \email{sipan.aslan@kaust.edu.sa}
\affiliation{ 
Statistics Program, King Abdullah University of Science and Technology (KAUST) 
}%


\date{\today}

\begin{abstract}
Causal interactions in time series networks can be dynamic and nonlinear, making it difficult to identify them using conventional linear causality estimations. We propose a novel approach, called Threshold Autoregressive Modeling for Causality (TAR4C), a causality detection approach built on threshold autoregressive (TAR) models, where a potential driver (cause variable)  acts both as a predictor and as a trigger (switching threshold) that governs which autoregressive process the target (effect variable) follows. Threshold nonlinearity is conceptualized here to determine causality. The flow of the target is forced to transition between regimes with distinct dynamics when the driver exceeds a data-driven threshold in the past. We propose a two‑stage inference procedure: Stage 1 tests for threshold connectivity (TC); Stage 2, conditional on a detected threshold effect, estimates threshold Granger causality (TGC). TAR4C is applied to a multichannel EEG dataset collected from a motor imagery and execution experiment. Delay-dependent directional interactions are observed among channels across different sites of the EEG map. The real-world application demonstrates the usefulness of the proposed approach for determining nonlinear causal connectivity in complex time-series networks, such as brain circuitry. The proposed model-based methodology extends to other complex networks of time series.


\end{abstract}
\maketitle
\begin{quotation}
Exploring connectivity in time-series networks enhances knowledge about underlying dynamics. Causal relations in such networks are often subject to regime shifts (e.g., state transitions) and asymmetric influences. We propose Threshold Autoregressive Modeling for Causality (TAR4C), an interpretable, state‑aware, nonlinear causality framework for causal discovery in time‑series networks. Inference proceeds in two steps. First, we detect threshold-nonlinearity-based connectivity, which shows a driver-controlled switching mechanism. Second, we test threshold Granger causality to evaluate the predictive power of the driver within the identified regimes. This procedure maps statistical associations that illustrate direction, delay, and strength, while accounting for asymmetric, discontinuous, and amplitude-dependent links. Estimation uses arranged autoregressions with heteroskedasticity‑robust tests and parsimonious model selection, and the procedure scales to network settings by attenuating confounding background activity before pairwise analysis. The EEG case study illustrates its practical application. The framework is general and can be extended to various multivariate signal systems, including those in economics, ecology, climate, and engineering.
\end{quotation}

\section{Introduction}
\label{sec:intro}
Time series networks are graphs in which each node is a source of a time series (or signal), and edges encode dynamic relationships -such as statistical dependence or directional (causal) influence- between the series. Addressing nonlinear characteristics of causal interactions in a time series network is important to understand underlying dynamics of its connectivity. This is especially crucial in neuroscience, where the connectivity of the brain circuitry defined as highly complex and varies with task demands and internal states \citep{biswas2022statistical}. Evidence accumulated over decades shows that brain connectivity has complex dynamics, which linear models only capture limited features. The complexity of brain connectivity motivates methods that treat causality as state‑dependent and time‑varying rather than static and linear \citep{sporns2016networks}. We introduce a new methodology and demonstrate its application in an EEG study of motor imagery (MI) and motor execution (ME) to show how it functions in real-world scenarios.

This paper proposes Threshold Autoregressive Modeling for Causality (TAR4C), a two-stage approach for detecting and characterizing nonlinear, regime-dependent causal interactions. In TAR4C, a potential driver (cause variable) serves both as a predictor and as a trigger for the regime-switching (threshold) that determines the autoregressive process governing the target (effect variable). Stage 1 tests for threshold connectivity (TC) by asking whether the driver induces a threshold effect on the target’s dynamics. Stage 2 estimates threshold Granger causality (TGC) by assessing whether past values of the driver improve prediction within the threshold‑structured model. The methodology is designed for network analysis: pairwise analysis between any selected pair of nodes is performed after accounting for the rest of the network. The EEG application demonstrates explored threshold-causal interactions in an interpretable way.

\subsection{Causal Connectivity in EEG networks}
A significant number of methods have been proposed as means of inferring functional and effective(i.e., causal) connectivity in EEG, with varying degrees of focus on specific objective aspects. These include Granger causality (GC) and related spectral analyses \citep{seth2015granger, Ombao2022}, Transfer Entropy (TE) \citep{vicente2011transfer, redondo2025measuring}, Directed Transfer Function (DTF) \citep{babiloni2005estimation}, Partial Directed Coherence (PDC) \citep{he2014nonlinear}, Bayesian time‑series modeling \citep{Prado2001, GORROSTIETA20123347, ZheYu2016, Chiang2017bvar}, and Machine / Deep Learning (ML / DL) modeling \citep{Tank20224267,chew2023granger}. 

Conventional GC relies on linear autoregressive models and evaluates whether including the past of a candidate driver reduces the prediction error of a target. While useful, the linear assumption limits inference when signals are governed by nonlinear dynamics \citep{shojaie2022granger}. To address this, alternative approaches extend GC to ML/DL models. These models can detect complex patterns and achieve strong predictive performance in brain imaging studies, yet they are often criticized for limited interpretability and voluminous data requirements. Ground truth labels at the scale needed for training are not always available in complex phenomena such as brain dynamics, and black‑box routines provide limited insight into mechanisms \citep{wein2021brain}. Such concerns suggest that developing an interpretable, explicitly nonlinear, and state-aware methodology would enable valuable knowledge about the underlying dynamics of causal connectivity.

\subsection{Challenges in EEG‑Based Causal Inference}
Inferring causal interactions in brain activity is challenging because the underlying dynamics are complex, nonlinear, and time‑varying. Neural processes -ranging from synaptic plasticity to oscillatory rhythms- shape cortical function and are rarely comprehended by simple linear models; for example, large‑scale oscillations may arise from nonlinear attractors, and chaotic dynamics can occur in neural circuits \citep{breakspear2017dynamic}. According to two influential frameworks aimed at explaining brain dynamics, state transitions are prominent features of neural dynamics. The metastable brain principle suggests that the brain shifts among semi‑stable coordination patterns \citep{tognoli2014metastable}, whereas the critical brain hypothesis suggests that neural activity operates near a phase transition \citep{beggs2003neuronal,beggs2008criticality}. To investigate nonlinear causality in time‑series networks such as EEG, we present TAR4C, a threshold autoregressive approach in which connectivity is associated with a regime-switching mechanism and explored by modeling its dynamics. The proposed method's cutting-edge feature threshold-governed regime shifting mechanism echoes switching in metastability and tipping‑point reconfigurations near criticality; however, our investigation does not assume either mechanism a priori and remains agnostic to these worth-mentioning hypotheses \citep{tognoli2014metastable,breakspear2017dynamic,beggs2008criticality}.

Additional challenges in inferring causality from EEG recordings reside in the limited capacity of EEG machinery and the physics of field propagation. Scalp EEG has millisecond temporal resolution but limited spatial resolution; each channel reflects a weighted mixture of several neural sources. Volume conduction can produce zero-lag correlations among nearby channels that do not reflect inter‑regional interactions. Preprocessing decisions (e.g., filtering, down-sampling, referencing) and model settings also affect estimated temporal dependencies and can introduce phase distortions or spectral leakage. Approaches such as source localization and independent component analysis (ICA) can reduce these effects, but they rely on additional assumptions and none remove them entirely \citep{bastos2016tutorial}. Accordingly, causal relations inferred in the EEG setting should be comprehended as model‑based effective connectivity between signals rather than direct synaptic transmission, and findings should be interpreted cautiously regarding neural specificity \citep{bastos2016tutorial}. EEG-based analyses can still suggest directionality of information flow, yet -given these limitations- they provide approximate indicators of underlying connectivity \citep{bastos2016tutorial, Ombao2022}.

\subsection{Objectives and Contributions}
The goal is to present a methodology, TAR4C, for detecting and characterizing nonlinear, time‑varying causal interactions in time‑series networks, with EEG used as an illustrative domain. The specific objectives are:

\textbf{Introducing threshold nonlinearity for connectivity inference.} TAR4C treats the driver’s state as a trigger that selects the regime governing the target. In this view, the threshold variable is a control signal: when the driver crosses a data‑driven threshold at a delay, the target switches to a different regime. Stage~1 tests whether such threshold‑induced regime switching exists, thereby identifying TC. The perspective of the study aligns with hypotheses that view oscillations among different states as a fundamental aspect of brain dynamics.

\textbf{Modeling nonlinear and time‑varying causal interactions.} Conditional on a detected threshold effect, TAR4C estimates a threshold‑structured autoregression in which the driver contributes as a predictor within each regime. Stage~2 evaluates TGC using heteroskedasticity‑robust tests, yielding a direction‑specific, delay‑specific measure of influence. Because regimes are selected by the driver’s state, the approach naturally encodes state dependence and time variation in connectivity.

\textbf{Comparing effective connectivity patterns in motor execution and imagery.} The methodology is demonstrated on multichannel EEG from MI and ME, focusing on prefrontal–motor interactions during bilateral gestures. The analysis asks whether directional influence differs across imagery and execution tasks. The EEG study is a well-designed real-world application that shows how the proposed method brings new perspectives into effective connectivity analyses.

\section{Exploring Causal Connectivity using TAR Models}
\label{sec:meth}
This section outlines the proposed TAR4C strategy for detecting regime‑shift related causality in time‑series networks and gives the basics of the threshold autoregressive (TAR) model. We study node pairs (e.g., EEG channels) after attenuating the confounding effect of the remaining (background) nodes in the network. The background is summarized by a low‑dimensional set of dynamic principal‑component scores computed in the frequency domain and then partialled out from the two nodes of interest (NOI). On these confounding free time series, we estimate TAR models in which a candidate driver acts both as a predictor and as a threshold variable. This setup lets us (i) test for threshold connectivity (existence of a threshold effect) and, if present, (ii) assess threshold Granger causality (predictive influence within regimes).

\subsection{Inference in high-dimensional time series networks}
High‑dimensional networks, such as multichannel EEG, contain many interacting nodes, which can confound pairwise inference. When the goal is set to discover the directed interaction between two specific nodes, it is practical to first remove the global background activity in a principled way and then analyze the residual interaction. We adopt a spectral dynamic principal‑component (sDPCA) approach for this purpose.

Let, $N=\{Y_t^{*},X_t^{*},Z_{1,t},\ldots,Z_{n-2,t}\}$, denote an $n$-node network. Here $Y^{*}$ and $X^{*}$ are the two nodes of interest; $\mathbf Z_t=(Z_{1,t},\ldots,Z_{n-2,t})^\top$ collects the remaining nodes. Define $H(\cdot)$ to be an sDPCA filter that maps $\mathbf Z_t$ to a small vector of dynamic scores $\mathbf p_t$ capturing the main frequency‑specific modes of the background:
$$
\mathbf p_t = H(\mathbf Z_t)\in\mathbb R^{k},\qquad k\ll n.
$$
We then work with "net" (interference‑reduced) series obtained by projecting $Y_t^{*}$ and $X_t^{*}$ on the span of $\mathbf p_t$ and taking residuals:
$$
Y_t = \tilde Y_t = Y_t^{*}-\widehat{\mathbb E}\big[Y_t^{*}\mid \mathbf p_t\big], 
\qquad
X_t = \tilde X_t = X_t^{*}-\widehat{\mathbb E}\big[X_t^{*}\mid \mathbf p_t\big].
$$
In practice, $\widehat{\mathbb E}(\cdot\mid \mathbf p_t)$ is a fitted regression (linear or semiparametric) on $\mathbf p_t$; $\mathbf p_t$ may include the leading dynamic principal‑component scores. This yields low‑variance, background‑adjusted series $(Y_t,X_t)$ on which we conduct the two‑stage TAR4C inference (threshold effect test, then threshold‑Granger test). This reduction‑and‑residualization step is computationally straightforward, preserves temporal structure, and has been shown to work satisfactorily for high‑dimensional time series \citep{brillinger2001time,aslan2024practical}.

\subsection{Background on the threshold autoregressive model}
The concept of separate regimes based on a change point was first introduced in linear regression by \citet{quandt1958estimation} and was extended to nonlinear time‑series analysis through threshold nonlinearity by \citet{tong1978threshold} and \citet{tong1980threshold}. The TAR model builds on the idea of representing systems whose dynamics change when a threshold variable crosses specific values. TAR modeling is able to capture and generate key nonlinear features, such as limit cycles, abrupt shifts, and asymmetries, by combining autoregressive components within regimes. This piecewise‑linear structure eases estimation and interpretation and is well-suited to signals with bursts and regime shifts. Comprehensive overviews appear in \citet{hansen2011threshold} and \citet{chen2011review}. Here, we use TAR as a foundation for the proposed procedure for causal inference.

A stochastic process $Y_{t}$ is TAR with $k$ regimes if it can be represented as follows
\begin{equation}\label{eq1}
       Y_{t}  =\sum_{j=1}^{k} \left [ (\phi_{0}^{(j)} + \sum_{i=1}^{p_j}\phi_{i}^{(j)}Y_{t-i}) \ \mathbb{I}(r_{j-1}<Z_{t-d}\leq r_{j}) \right ] + \ \varepsilon_{t},
\end{equation}
where $j=1,\ldots,k$; $Z_{t-d}$ is the threshold variable with positive integer delay parameter $d$; $p_j$ is for $j^{th}$ regime AR lag order; $\phi_{0}^{(j)}$ is intercept and $\{\phi_{i}^{(j)}|i=1,2,\ldots, p_j \}$
are AR parameters for $j^{th}$ regime; $\mathbb{I}(.)$ is indicator function ({$\mathbb{I}(\mathcal{A})=1$ if $\mathcal{A}$ is true, else $\mathbb{I}(\mathcal{A})=0$}) and $r=(r_{1}, \ldots, r_{k-1})$ satisfying $r_{1}< \ldots <r_{k-1}$ are the threshold values and lie in the bounded subset $[min\left\{Z\right\},max\left\{Z\right\}]$ of threshold variable sample space; $r_{0}=-\infty$ and $r_{k}=\infty$;  regression error term, 
$\{\varepsilon_{t}\}$ is a sequence of martingale differences satisfying $\ex(\varepsilon_{t}|F_{t-1}) = 0$,  ${sup}_{t}\ex( |\varepsilon_{t}|^{\delta}|F_{t-1}) <\infty$ $almost$ $  surely$ for some $\delta>2$, where $F_{t-1}$ is the $\sigma$-field generated by $\{\varepsilon_{t-i}|i=1,2\ldots\}$.
For example, if at time $t$, let $Z_{t-d} = a \in (r_{j-1}, r_{j})$ then the active regime at that time is characterized by linear AR model;
\[
Y_t = \phi_{0}^{(j)}+ \sum_{i=1}^{p_j}\phi_{i}^{(j)}Y_{t-i} \ + \ \varepsilon_{t}^{(j)}.
\]
The overall process $Y_t$ is nonlinear whenever at least two regimes differ, even though each regime is linear.

The unknown parameters for the model in Equation (\ref{eq1}) can be assembled as follows
\[ \Omega= \left [ (\phi_{0}^{(1)},\phi_{i:p_1}^{(1)}), \ldots,(\phi_{0}^{(k)},\phi_{i:p_k}^{(k)})\right ].\]
The hyperparameters of the model can be shown as;
\[\Gamma= \{k,p_{j}, d, r\in[\text{min}(Z_{t-d}),\text{max}(Z_{t-d})]\}.\]
Conditional on $\Gamma$, the model is piecewise linear, so least squares (LS) estimation is straightforward and, under i.i.d. Gaussian errors $\varepsilon_{t} \sim N(0,\sigma_{\varepsilon_{t}}^2)$, coincides with maximum likelihood. Conditional least squares (CLS) estimation is useful because it enables the search over threshold values and hyperparameters without imposing restrictions. Consistency of CLS holds under mild conditions such as stationarity and a discontinuous conditional mean function \citep{chan1993consistency}.

The LS estimators $\widehat{\Omega}$, $\widehat{\Gamma}$ jointly minimize the following objective function:
\begin{equation}\label{eq2}
\mathcal{S}({\Omega},{\Gamma})=\sum_{t=1}^{T} \left [ Y_{t}-\sum_{j=1}^{k}\left [(\phi_{0}^{(j)}+ \sum_{i=1}^{p_j}\phi_{i}^{(j)}Y_{t-i})\mathbb{I}(r_{j-1}<Z_{t-d}\leq r_{j}) \right ] \right ]^2,
\end{equation}
\begin{equation}\label{eq3}
\widehat{\Omega},\widehat{\Gamma}=\argminC_{\Omega,\Gamma} \mathcal{S}({\Omega},{\Gamma})
\end{equation}
where $T$ denotes the sample size. The LS estimators $\widehat{\Omega}$ and the minimization of the objective function in Equation (\ref{eq2}) can be achieved conditionally by a grid search considering all viable combinations of hyperparameters $\{j,p_{j},r,d|j=1, \ldots, k\}$. The optimization requires the execution of approximately $T^{k-1}/(k-1)! \times d\times p_{1}\times \ldots \times p_{k}$ arranged autoregressions. Arranged autoregression is an autoregression where cases are rearranged by $\Gamma$-dependent sample splitting. Note that rearrangement of the sample based on the threshold values does not change the temporal structure of the series within cases \citep{tsay1989testing}. Asymptotic theory for LS/CLS estimation and inference in TAR models is developed by \citet{chan1993consistency} and \citet{zhang2023least} (among others).

\noindent \underline{Remark.} The model in Equation~(\ref{eq1}) is called {\it self-exciting} threshold autoregressive (SETAR) model if the threshold variable equals the series itself, i.e., $Z_{t-d}=Y_{t-d}$. In that case, the regime of the time series $Y_{t}$ would be determined by its own past value, $Y_{t-d}$.
\subsection{Proposed Approach: TAR for causality}
\label{proposed}
This section proposes a 2-regime TAR model-based approach to determine threshold causal interactions between a given pair of time series residing in a network. The main objective is to map the net causal influence for a pair of nodes after mitigating interference from outer nodes. Let \( N = \{Y^*_t, X^*_t, Z_{1,t}, \ldots, Z_{n-2,t}\} \) represent an \( n \)-dimensional network of time series; \( Y^* \) and \( X^* \) are the nodes of interest between which we aim to discover the causal relationship. Considering \( Y \) and \( X \) as the interference-free counterparts of \( Y^* \) and \( X^* \), for which the impact of outer nodes is removed, the "net" activities at time \( t \) are denoted as \( Y_t \) and \( X_t \). Thereby, the 2-regime TAR model for the node $Y$, and the node $X$, which is treated as both the threshold and predictive variable, can be formulated as follows: 
\begin{equation}\label{eq4}
 Y_{t} = \begin{cases}
 \phi_{0}^{(1)}+ \sum_{i=1}^{p_1}\phi_{i}^{(1)}Y_{t-i} + \sum_{\ell=1}^{q_1}\theta_{\ell}^{(1)}X_{t-\ell} \ + \ \varepsilon_{t}, & \text{if } u(X_{t-d}) \leq r \\
 \phi_{0}^{(2)}+ \sum_{i=1}^{p_2}\phi_{i}^{(1)}Y_{t-i} + \sum_{\ell=1}^{q_2}\theta_{\ell}^{(2)}X_{t-\ell} \ + \ \varepsilon_{t}, & \text{if } u(X_{t-d}) > r
\end{cases}, 
\end{equation}
where $u(X_{t-d})$ is the function constructs the threshold variable; $d$ denotes delay parameter with positive integer that establishes the time delay at which the threshold function $u(X_{t-d})$ significantly influences the structure of $Y_t$; \(r\in[\text{min }u(X_{t-d}), \text{max }u(X_{t-d})]\) denotes threshold value; $p_1$ and $p_2$ indicates $1^{th}$ and $2^{nd}$ regime lag order of the node $Y_t$, respectively;  $q_1$ and $q_2$ indicates $1^{th}$ and $2^{nd}$ regime lag order of the node $X_t$, respectively; $\{\phi_{0}^{(1)}; \phi_{i}^{(1)}|i=1,2,\ldots, p_1 ; \theta_{\ell}^{(1)}|\ell=1,2,\ldots, q_1 \}$ are parameters for $1^{th}$ regime; and $\{\phi_{0}^{(2)}; \phi_{i}^{(2)}|i=1,2,\ldots, p_2 ; \theta_{\ell}^{(2)}|\ell=1,2,\ldots, q_2 \}$ are parameters for $2^{nd}$ regime; $\{\varepsilon_{t}\}$ is a sequence of martingale differences satisfying $\ex(\varepsilon_{t}|F_{t-1}) = 0$,   ${sup}_{t}\ex( |\varepsilon_{t}|^{\delta}|F_{t-1}) <\infty$ $almost$ $  surely$ for some $\delta>2$, where $F_{t-1}$ is the $\sigma$-field generated by $\{\varepsilon_{t-i}|i=1,2\ldots\}$. 

The unknown set of parameters for the model in Equation (\ref{eq4}) can be shown as follows:
\begin{equation}\label{eq5}
\Omega = \left\{ \Phi, \Theta \right\}, \quad \Phi = \left\{ \phi_0^{(1)}, \phi_{i:p_1}^{(1)}, \phi_0^{(2)}, \phi_{i:p_2}^{(2)} \right\}, \quad \Theta = \left\{ \theta_{\ell:q_1}^{(1)}, \theta_{\ell:q_2}^{(2)} \right\}.
\end{equation}
The hyperparameters of the model can be collected as \[\Gamma= \{p_{1},p_{2},q_{1},q_{2}, d, r\in[\text{min }u(X_{t-d}),\text{max }u(X_{t-d})]\}.\] The parameters in (\ref{eq4}) can be estimated using the LS method conditional on these hyperparameters, similar to minimizing the objective function shown in (\ref{eq2}) and (\ref{eq3}). To discover TAR model-based causal interactions in a network \( N \) of multiple nodes, we specify the following definitions related to (\ref{eq4}) and (\ref{eq5}).
\begin{widetext}
\begin{definition}\textnormal{\textbf{:}} 
\text{ } \\
\( \text{ }\forall (X,Y) \in N, X \xrightarrow[\text{Threshold}]{\text{Connected}} Y \text{ if }\) 
\(\exists \text{ }{\Phi} \neq \{\emptyset\}\text{ } \land \text{ }{\Theta} = \{\emptyset\}\text{ } \land \text{ }r\in[\underline{u(X_{t-d})},\overline{u(X_{t-d})} ]   \) \end{definition}
\noindent{\textbf{(X is threshold-connected to Y if X exerts a threshold nonlinearity effect on Y)}}
\end{widetext}
The first definition implies that the presence of threshold nonlinearity in the TAR model (\ref{eq4}) suffices to show that \( X_t \) has a threshold causal effect on \( Y_t \). In this case, the parameter set \( \Theta \) is empty. That is, even if the past values of \( X_t \) do not have a significant predictive effect, the threshold functional of \( X_{t-d} \) still influences the structure of \( Y_t \) by controlling which regime \( Y_t \) follows. To explore the TC in Definition 1, we utilize the threshold nonlinearity test scheme suggested by \cite{hansen1996inference}.

The model given in (\ref{eq4}) can also be rewritten to illustrate Hansen's test in a simpler way. Let $W_{t}= (1\text{ }Y_{t-1}\text{ }\ldots\text{ }Y_{t-p}\text{ }X_{t-1}\text{ }\ldots\text{ }X_{t-q} )^\prime $ be a $m\times1$ vector where $p_1=p_2=p$; $q_1=q_2=q\text{ }\text{ }\text{ }\text{ }$ and\text{ }\text{ }\text{ } $m=p+q+1$;\text{ }\text{ }\text{ }\text{ } $\pi_{1}= (\phi_{0}^{(1)}\text{ }\phi_{1}^{(1)}\text{ }\ldots\text{ }\phi_{p}^{(1)}\text{ }\theta_{1}^{(1)}\text{ }\ldots\text{ }\theta_{q}^{(1)} )^\prime $\text{ } and\text{ } $\pi_{2}= (\phi_{0}^{(2)}\text{ }\phi_{1}^{(2)}\text{ }\ldots\text{ }\phi_{p}^{(2)}\text{ }\theta_{1}^{(2)}\text{ }\ldots\text{ }\theta_{q}^{(2)} )^\prime $ then the 2 regime TAR model in (\ref{eq4}) takes the form;
\begin{flalign}\label{eq6}
	& \text{M2:}
	& Y_t= \pi_{1}^\prime W_{t}\mathbb{I}(u(X_{t-d})\leq r) + \pi_{2}^\prime W_{t}\mathbb{I}(u(X_{t-d})> r) + \varepsilon_t.\phantom{AAAAAA}
\end{flalign}
In the absence of threshold nonlinearity, $\pi_1=\pi_2$   (i.e., $r = r_0$) the linear autoregressive distributed lag (ADL) model can be expressed as follows:
\begin{flalign}\label{eq7}
	& \text{M1:}
	& Y_t= \pi_{1}^\prime W_{t} + \varepsilon_t.\phantom{AAAAAAAAAAAAAAAAAAAAAAAAAAAA}
\end{flalign}
The unknown parameters of the models represented in (\ref{eq6}) and (\ref{eq7}) can be shown as \( \Pi_\text{\tiny M2} = [\pi_{1}, \text{ }\pi_{2}, \text{ }r, \text{ }d ]\) and \( \Pi_\text{\tiny M1} = [\pi_{1}]\), respectively. Under the assumption $\ex(\varepsilon_{t}|F_{t-1}) = 0$ the appropriate estimation method is LS method for both. The LS estimator of $\widehat{\Pi}_\text{\tiny M1}$ and $\widehat{\Pi}_\text{\tiny M2}$, which is conditional on fixed hyperparameters $\{r, d\}$, can be found by the minimization of following objective functions;
\begin{align}\label{eq8}
	& \widehat{\Pi}_\text{\tiny M1} = \argminC_{\Pi_\text{M1}} \sum_{t=1}^{T}\big[Y_t - \pi_{1}^\prime W_{t} \big]^2
\end{align}
\begin{align}\label{eq9}
	& \widehat{\Pi}_\text{\tiny M2} = \argminC_{\Pi_\text{M2}} \sum_{t=1}^{T}\big[Y_t - \pi_{1}^\prime W_{t}\mathbb{I}(u(X_{t-d})\leq r) - \pi_{2}^\prime W_{t}\mathbb{I}(u(X_{t-d})> r) \big]^2
\end{align}
Models given in (\ref{eq6}) and (\ref{eq7}), which are the most restricted, have a nested structure. Hence, comparing them using a model selection criterion is equivalent to a likelihood ratio test \citep{hansen1999testing}. 

Let $S_\text{\tiny M1}={\hat{\varepsilon}_\text{\tiny M1}}^\prime\times{\hat{\varepsilon}_\text{\tiny M1}}^\text{}$ and
$S_\text{\tiny M2}=\hat{\varepsilon}^\prime_\text{\tiny M2}\times {\hat\varepsilon_\text{\tiny M2}}^\text{}$ denote the sum of squared residuals obtained by LS estimation in (\ref{eq8}) and (\ref{eq9}), respectively; and let $r_0$ is denoting the true value of the threshold value $r$; then testing the existence of threshold non-linearity can be accomplished by the testing the hypothesis of $\text{H}_0: r = r_0$ considering the following likelihood ratio (LR) test statistic;
\begin{equation}\label{eq10}
	LR(r)=T\bigg(\frac{S_\text{\tiny M1}-S_\text{\tiny M2}(\hat r)}{S_\text{\tiny M2}(\hat r)}\bigg).
\end{equation}
The test of \( H_0 \) (i.e., \text{M1} against \text{M2}, linear vs. nonlinear model) involves rejecting large values of \( LR(r_0) \). This is the likelihood ratio test under the assumption that \( \varepsilon_t \) are independent \( N(0,\sigma^2) \), corresponding to the conventional F (Wald) and Lagrange multiplier (score) tests. However, since the nuisance parameter \( r \) is not identified under the null hypothesis of no threshold effect, the asymptotic distribution of \( LR \) does not follow a chi-square distribution. To overcome this, \cite{hansen1996inference} demonstrates that $p$-values can be approximated using residual bootstrap simulation. The test procedure also provides a heteroskedasticity-consistent Lagrange multiplier test that adjusts for heteroskedastic error terms. The asymptotic distribution of the threshold nonlinearity test and some of the major theoretical problems of inference for TAR model estimation are comprehensively studied by \cite{hansen1996inference, hansen1997inference, hansen1999testing,hansen2000sample}. Through testing the existence of threshold non-linearity, we aim to discern the connectivity between time series, shedding light on the causal connectivity.

\begin{widetext}
\centering
\begin{definition}
\textnormal{\textbf{:}} \text{ } \\ \(\text{ }\forall (X,Y) \in N, X \xrightarrow[\text{Threshold G}]{\text{Cause}} Y \text{ if }\) \(\exists \text{ }{{\Theta}} \neq \{\emptyset\}\text{ } \land \text{ }r\in[\underline{u(X_{t-d})},\overline{u(X_{t-d})} ] \)  
\end{definition}
\textbf{(X threshold Granger causes Y if X has both threshold nonlinearity effect on Y and also having predictive power on Y)}.
\end{widetext}

The second definition specifies the TGC by incorporating the impact of the threshold node's predictive ability in the model. Once threshold nonlinearity is verified, causal interactions between nodes can be explored using the TAR model by including the threshold node \( X_{t-d} \) as a predictor of \( Y_t \). Identifying TGC requires statistically significant coefficients for the time-lagged terms of \( X_t \). To avoid over-parameterization in representing causal interactions, we propose a model selection criterion using the heteroscedasticity-robust Wald statistics from \cite{hansen1997inference} as a penalty term. The Wald test assesses the predictive capacity of the threshold variable as an explanatory covariate. It should be noted that \cite{li2006testing} was the first to mention GC in the presence of threshold effects and used the TGC nomenclature. However, \cite{li2006testing} utilized the dependent variable itself as the threshold variable. Although we adopt the TGC terminology, our proposed methodology differs in terms of the causal connectivity framework.

Considering the model given in (\ref{eq4}) the hypothesis to be tested for TGC can be expressed as follows: 
\begin{equation}\label{eq11}
\text{H}_0: \theta_{1}^{(1)}=\theta_{1}^{(2)}=\ldots=\theta_{q_1}^{(1)}=\theta_{q_2}^{(2)}=0,
\end{equation}
where the alternative hypothesis discerns the TGC by stating that at least one coefficient is nonzero. If the threshold value $r$ splitting the sample and delay $d$ are known, the pointwise heteroscedasticity-consistent Wald statistic can be shown as follows:
\begin{widetext}
\begin{equation}\label{eq12}
 \Delta_T(r,d) = \bigg(\text{C}\hat\Omega(r,d)\bigg)^\prime \bigg[\text{C}\bigg(M(r,d)^{-1}V(r,d)M(r,d)^{-1}\bigg)\text{C}^\prime\bigg]^{-1}\bigg(\text{C}\hat\Omega(r,d)\bigg),
\end{equation}
\end{widetext}
where $\text{C}$ is the selector matrix;\newline  $M(r,d)=\sum_{t=1}^{T}W_t(r,d)W_t(r,d)^\prime$;\newline
\(V(r,d)=\sum_{t=1}^{T}W_t(r,d)W_t(r,d)^\prime \hat{\varepsilon_t}^2\);\newline  $W_t(r,d) = \bigg(W_t^\prime \mathbb{I}(u(X_{t-d})\leq r)\text{ }\text{ } W_t^\prime \mathbb{I}(u(X_{t-d})> r)\bigg)^\prime$, then according \citet{hansen1996inference, davies1987hypothesis} the appropriate testing of $\text{H}_0$ can be done by using following LR statistic

\begin{equation}\label{eq13}
 \Delta_T = \sup_{(r,d)\in \Gamma } \Delta_T(r,d).
\end{equation}

Unlike the test statistic in (\ref{eq10}), this test treats \( (r,d) \) as fixed under both the null and alternative hypotheses, resulting in a standard testing procedure where the asymptotic distribution of \( \Delta_T \) is \( \chi^2_{\rho(C)} \), with \( \rho(C) \) being the number of restrictions. The test rejects the null of the restricted model for large values of \( \Delta_T \). Note that the restricted model tested in (\ref{eq11}) is not the only hypothesis implying TGC; all alternative restricted models involving \( \Theta \) imply TGC. Therefore, restricted models indicating TGC should be explored through a grid search, as \( \Delta_T \) varies over hyperparameters \( p_1, p_2, q_1, q_2 \). Moreover, \( \Delta_T \) measures the deviance between restricted and unrestricted models, making it suitable for use in model selection criteria to choose a simpler, statistically valid model.

Thus, after confirming threshold nonlinearity, to obtain a parsimonious representation of the causal interaction between \( X \) and \( Y \), we propose a model selection criterion, a modified Schwarz criterion (SC):

\begin{equation}\label{eq14}
    {mSC}_{T} = \Delta_T - 2 \log(T) K,
\end{equation}

where \( K \) is the number of parameters in the model. Estimating TGC then involves maximizing the objective function in (\ref{eq14}):

\begin{equation}\label{eq15}
    \max_{\Gamma} \ {mSC}_{T}.
\end{equation}

Maximizing (\ref{eq14}) provides estimates of parameters \( \hat{\Omega} \) and \( \hat{\Gamma} \). Similar model selection methods and related remarks have been discussed by Raftery (1995), Kass and Raftery (1995), Liu et al. (1997), and Gonzalo and Pitarakis (2002). The complete TAR4C procedure is outlined in Procedure \ref{proc:TAR4C}.

\IncMargin{1.0em}
\begin{algorithm}[ht!]
    \SetKwData{Left}{left}
    \SetKwData{This}{this}
    \SetKwData{Up}{up}
    \SetKwFunction{Union}{Union}
    \SetKwFunction{FindCompress}{FindCompress}
    \SetKwInOut{Input}{Input}
    \SetKwInOut{Output}{Output}
    \SetKwInOut{Data}{Data}
    \SetKwInOut{Parameters}{Parameters}
    \SetKwInOut{Scope}{Scope}
    \SetKwInOut{Network}{Network}
    \SetKwInOut{Decision}{Decision}

    \small{
    \setstretch{0.9} 
    \BlankLine
    \Network{\( N = \{Y^*_t, X^*_t, Z_{1,t}, \ldots, Z_{n-2,t}\}  \)}
    \BlankLine
    \Scope{$ \text{Analyze} \ {\text{TC}}_{X\to Y} \ \text{and} \ {\text{TGC}}_{X\to Y} $}
    \BlankLine
    \Input{$X, Y \in N$ (nodes of interest in a network, $N$)}
    \BlankLine
    \Parameters{$\Omega, \Gamma$}
    \BlankLine
    \Data{$\{({X}_{t}, {Y}_{t})\}_{t=1}^{T}$ and $u({X}_{t-d})$}
    \BlankLine 
    \Decision{(a) ${\text{TC}}_{X\to Y}$ \quad
            (b) ${\text{TGC}}_{X\to Y}$ \quad
            (c) ${\text{TC}}_{X \not\to Y}$} 
        \BlankLine}

\begin{framed}

\small{
    \setstretch{0.9} 
    \begin{enumerate}[label=\bfseries Step \arabic*.,leftmargin=*,labelindent=0em,itemsep=1pt,parsep=0pt,topsep=1pt]

        \item \textbf{Neutralize interference from other nodes:}
        \begin{enumerate}[label*=\bfseries \arabic*.,itemsep=1pt,parsep=0pt,topsep=1pt]
           \item \( X_t = X_t^{*}-\widehat{\mathbb E}\big[X_t^{*}\mid \mathbf p_t\big]\)
           \item \( X_t = X_t^{*}-\widehat{\mathbb E}\big[X_t^{*}\mid \mathbf p_t\big]\)
        \end{enumerate}

        \item \textbf{Testing threshold effect:}
        \begin{enumerate}[label*=\bfseries \arabic*.,itemsep=1pt,parsep=0pt,topsep=1pt]
            \item Specify the $u(X_{t-d})$ of node $X$ whose causative effect on $Y$ is of interest.
            \item Regarding the model in Equation (\ref{eq6}), test the null of linearity:
            \[ \text{H}_0: \pi_{1} = \pi_{2} \]
            \item Significant evidence against linearity indicates ${\text{TC}}_{X\to Y}$. If so, proceed to the next step.
            \item[] Fail to reject the null implies ${\text{TC}}_{X \not\to Y}$.
        \end{enumerate}

        \item \textbf{Identify causal representation by TAR:}
        \begin{enumerate}[label*=\bfseries \arabic*.,itemsep=1pt,parsep=0pt,topsep=1pt]
            \item \( \max_{\Gamma} \ {mSC}_{T} \)
            \item Obtain $\hat{\Omega}$ and $\hat{\Gamma}$ from the maximization.
            \item Conclude ${\text{TGC}}_{X\to Y}$ if $\hat{\Theta} \neq \{\emptyset\}$.
        \end{enumerate}

\end{enumerate}}

\end{framed}
\caption{TAR4C$(X,Y)$}\label{proc:TAR4C}
\end{algorithm}

\textbf{On the role of TAR models for exploring causal connectivity in a network:} Use of TAR models makes causality inference state‑dependent. When a driver crosses a data‑driven threshold at a delay, the target follows a different autoregressive regime. This mechanism exposes nonlinear and asymmetric influences that linear Granger tests ignore. Threshold tests localize the switching regions and reveal when and how the relation between two nodes changes. Within TAR4C, we first test for a threshold effect (TC). If detected, we then assess regime‑specific prediction from the driver to the target (TGC). The approach has costs. A grid search over threshold $r$, delay $d$, and lag orders increases computation and raises the risk of over-parameterization. These risks are managed by using heteroskedasticity‑robust tests and applying a parsimonious model‑selection rule (e.g., the modified Schwarz criterion) after the threshold test. In practice, TAR4C remains tractable for pairwise analysis because background activity is first condensed by sDPCA and partialled out, leaving "net" series on which arranged autoregressions are fit. The result is an interpretable, regime‑aware representation of causal connectivity suited to high‑dimensional time series networks.

\section{Analysis of Motor Imagery vs Execution EEG Data}
\label{sec:application}
\subsection{Dataset and the experimental procedure}
This paper examines differences in brain network dynamics between motor imagery (MI) and execution (ME) tasks using 64-channel EEG recordings collected during an experiment, available at PhysioNet.org \citep{schalk2004bci2000}. Participants executed or imagined motor tasks based on visual cues, performing gestures like opening and closing (OC) both fists or feet (BHs/BF). EEG data were sampled at 160 Hz with annotations. Preprocessing removed environmental (e.g.,line freq, vibration), instrumentation (e.g., in surface electrodes), and biological (e.g., eye blinks, muscle activity) based artifacts by applying a high-pass filter at 0.5 Hz, removing 60 Hz line noise, and performing Independent Component Analysis (ICA) via MNE routines \citep{GramfortEtAl2013a}. Despite preprocessing, some issues (e.g., sawtooth traces and excessive blinking) remained in certain recordings. We focused our analysis on 35 right-handed subjects (17 females, 18 males) with clean recordings. To study causal connections, we selected four channels (AF3, AF4, C3, C4) over the prefrontal and motor cortices (Figure \ref{64-channel-EEG-placement}), regions that are likely to be active during movement and imagery. The explored connectivities are outlined in Table \ref{outline-table}.

\begin{table*}[ht]
    \centering
    \begin{tabular}{cc}
        \begin{minipage}{0.35\textwidth}
            \centering
            \includegraphics[width=\textwidth, trim=24cm 9.0cm 24cm 9.0cm, clip]{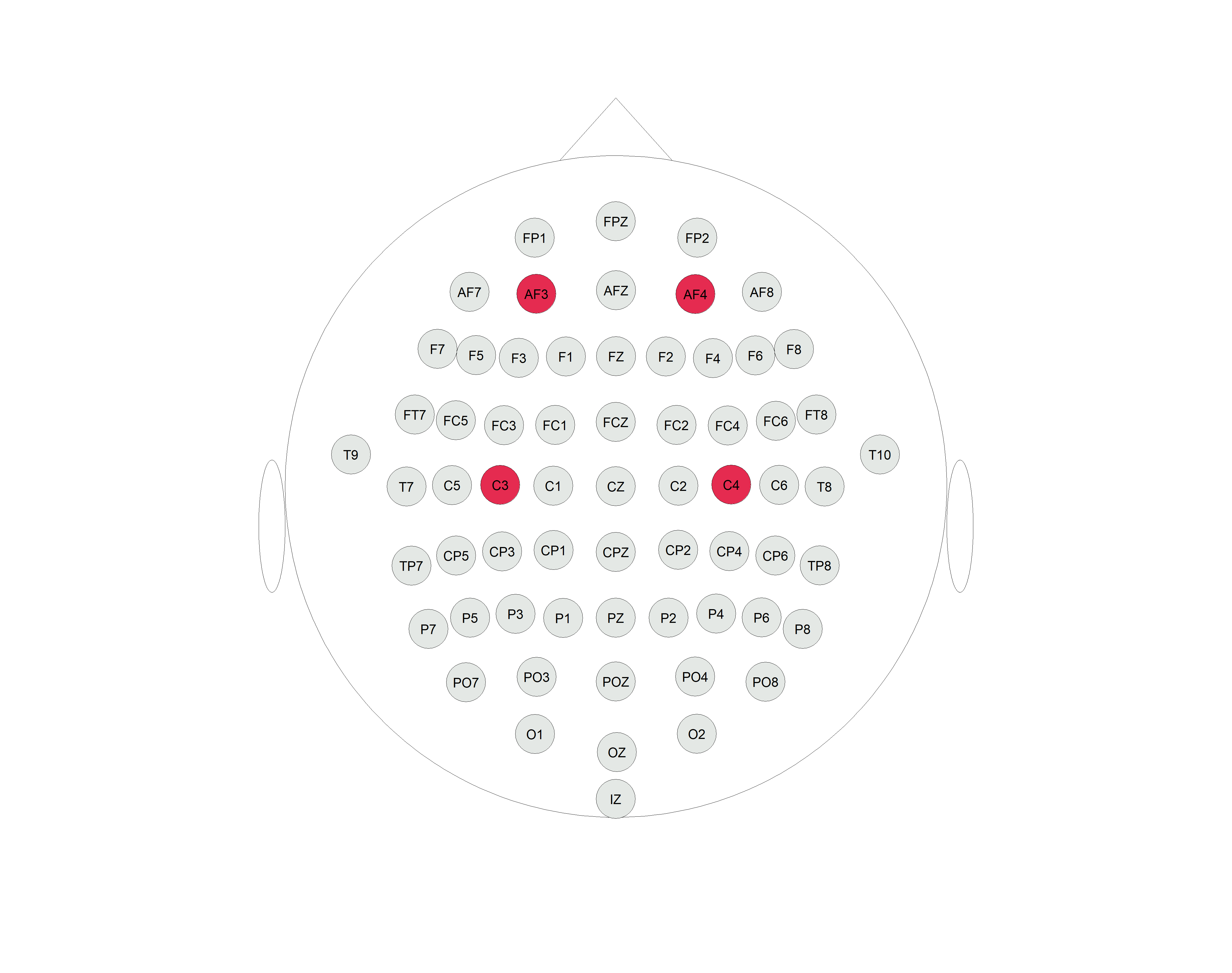}
            \captionof{figure}{Channels of interest.}
            \label{64-channel-EEG-placement}
        \end{minipage} 
        &
        \begin{minipage}{0.75\textwidth}
            \centering
            \caption{Illustration of links tested.}
            \label{outline-table}
            \begin{tabular}[t]{lcc}
                \toprule
                &Motor Execution&Motor Imagery\\
                \midrule
                \raisebox{2\height} {BHs}&
                \centering
                \includegraphics[width=0.25\textwidth, trim=34cm 40.0cm 34cm 25.0cm, clip]{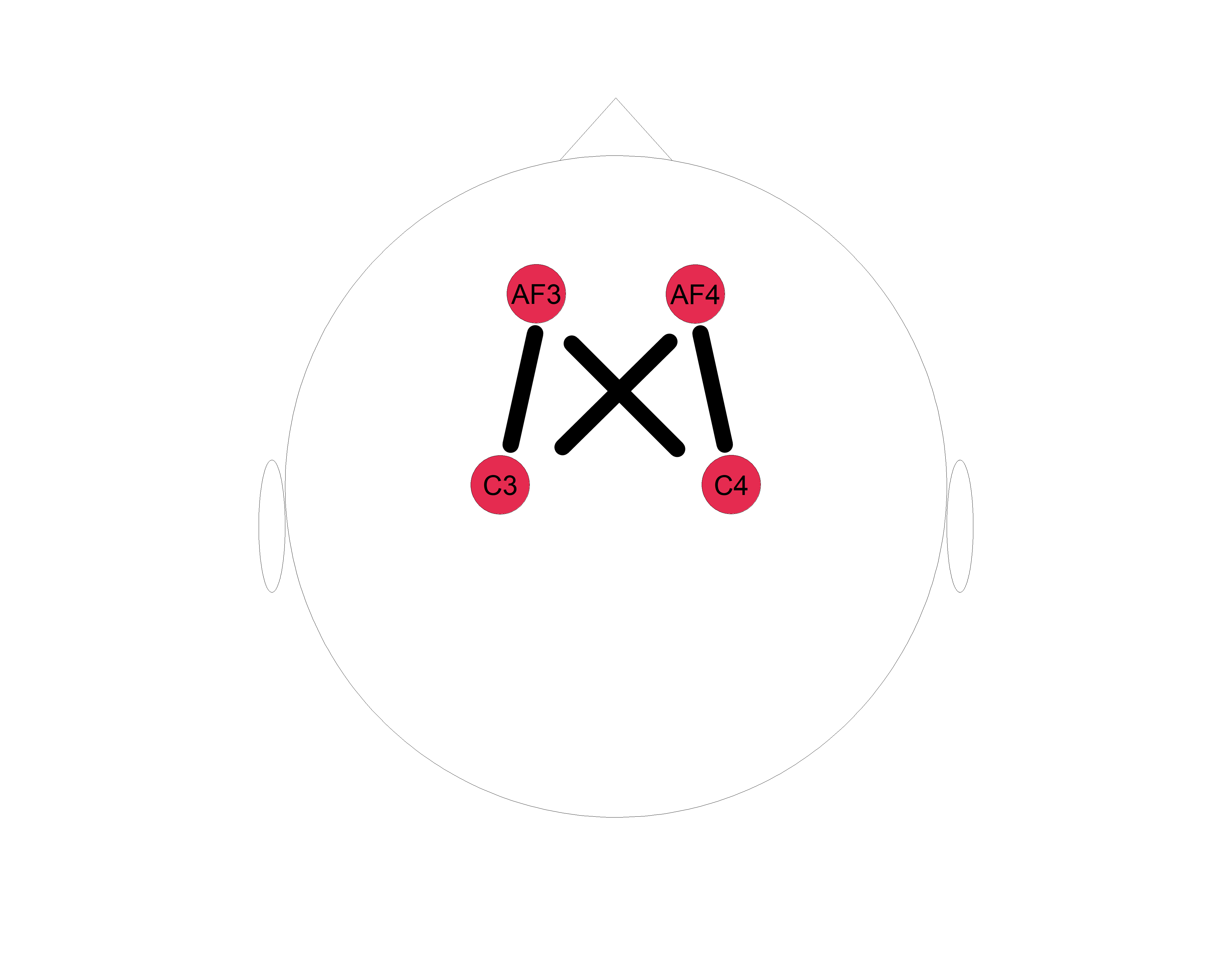} &
                \includegraphics[width=0.25\textwidth, trim=34cm 40.0cm 34cm 25.0cm, clip]{Figures/outline-arrow.png}  \\
                \raisebox{2\height} {BF}&
                \centering
                \includegraphics[width=0.25\textwidth, trim=34cm 40.0cm 34cm 25.0cm, clip]{Figures/outline-arrow.png} &
                \includegraphics[width=0.25\textwidth, trim=34cm 40.0cm 34cm 25.0cm, clip]{Figures/outline-arrow.png} \\
                \bottomrule
                ~~\\
                ~~\\
                ~~\\
            \end{tabular}
        \end{minipage}
    \end{tabular}
\end{table*}

\subsection{Analysis using the TAR4C model}
Each subject repeated the relevant task (i.e., task1: OC of BHs/BF, task2: imagery OC of BHs/BF) 3 times (i.e., runs). Within these three task-oriented runs, they performed each motor execution/imagery motion (i.e., BHs, BF) approximately 23 times in total. Each epoch contains about 4 seconds (i.e., 640 observations). For each task, 23 test results were combined following \cite{becker1994combining} to determine the presence of threshold non-linearity in the task for each subject. The method for inferring connectivity for a channel pair, called \textit{TAR4C}, is described in Chapter \ref{sec:meth}. In the application step, we focused on two index metrics to summarize connectivity results for all subjects.

\subsubsection{Threshold Connectivity Index}
The Threshold Connectivity Index (TCI) quantifies, considering all the subjects, condition-specific connectivity between two EEG channels based on the detection of threshold nonlinearity in their interaction. The TCI is calculated as follows:
\[
\text{TCI}_{XY} = \left(\frac{1}{12 S}\sum_{d=1}^{12} \mathbb{I}\left(s^{(d)} \ge \tau S \right) s^{(d)}\right)\times 100,
\]
where $s^{(d)}$ is the number of subjects out of the total number of subjects $S$ rejecting the linearity at delay $d$, $\tau$ is the group‑level prevalence rate. This indicates the number of subjects exhibiting threshold nonlinearity at the specified delay. \(\mathbb{I}\left(s^{(d)} \geq \tau S \right)\) is an indicator function; it equals 1 when $s^{(d)}\geq \tau  S$,  meaning that the number of subjects rejecting linearity at delay $d$ equals or exceeds $\tau $ (e.g., 70\%) of the total subjects. In practical terms, a higher TCI value suggests stronger threshold-based connectivity between the two channels, which is visualized in network diagrams with thicker lines representing stronger connections. The analysis procedure for calculating TCI involves the following steps:
\begin{table*}[ht]
	\centering
	\caption{ Outline of the implementation for TCI calculation.}
	\label{outline-analysis-table-TCI}  
    \small
	\begin{tabular}[t]{||l|c||}
		\toprule
		& {Motor Execution / Motor Imagery}\\
		\midrule
		\makecell{Gesture \\ to Perform}& 
		\begin{minipage}{4.5in}
        \raggedright
		For a given subject;
		Let $X$ and $Y$ represent two nodes in the EEG network.
		Compare whether $X$ is Threshold Connected to $Y$ using the proposed approach given in Section \ref{proposed}.
	\begin{enumerate}
    \setlength\itemsep{0em}
    \item Extract a total of approximately 23 epochs from the three task-oriented runs for the relevant gesture.
    \item Remove the influences of other nodes from the nodes of interest by regressing out the first two/three dynamic spectral principal components (e.g., the components can be desired to explain 85\% of the total variance at least).
    \item Define different functionals of the threshold variable \( X_{t-d} \), such as \( \lvert X_{t-d} \rvert \), \( \lvert X_{t-d} \rvert^2 \), etc., where \( d = 1, \ldots, 12 \). The results in the paper are the findings obtained when \( X_{t-d} \) is considered as a threshold variable.
    \item Apply the threshold nonlinearity test using the 12th lag TAR model for each epoch and each functional.
    \item Combine the test results over approximately 23 epochs for each functional and each delay value \( d \).
    \item Determine if the combined test result favors threshold nonlinearity. If so, note that \( X \) is the threshold connected to \( Y \) for the given subject and gesture.
    \item[$\blacksquare \phantom{.}$] Repeat all steps for each subject. If threshold nonlinearity is detected in at least 70\% of the subjects for each configuration tested, conclude that \( X \) has a threshold connectivity with \( Y \).
\end{enumerate}
		\end{minipage}  \\
		\bottomrule
	\end{tabular}
\end{table*}

\subsubsection{Threshold Granger Causality Index}
The Threshold Granger Causality Index (TGCI) extends the TCI by incorporating an additional step that tests the predictive power of the threshold variable. After establishing the presence of threshold nonlinearity, the TGCI assesses whether the past values of the threshold variable significantly predict the future values of the dependent variable using the Wald test. This index, therefore, provides a measure of directional causal influence between EEG channels. The TGCI is calculated as follows:
\[
\text{TGCI}_{XY} = \left(\frac{1}{12 S}\sum_{d=1}^{12} \mathbb{I}\left(s^{(d)} \ge \tau S \right) \text{Wald}^{(d)}\right)\times 100, 
\]
where $\text{Wald}^{(d)}$ is the number of subjects out of the total $S$ passing the Wald test in favor of the predictive power of the threshold variable $X$ at delay $d$.  This index quantifies the strength of the causal influence from $X$ $\to$ $Y$. The TGCI reflects the strength and directionality of the causal relationship between the channels and is used to visualize the directional flow of information in network diagrams. A higher TGCI value indicates a stronger causal influence from the threshold variable to the dependent variable.
The analysis procedure for calculating TGCI builds on the steps used for TCI and includes the following additional steps:

\begin{table*}[ht]
\centering
\caption{ Outline of the implementation for TGCI calculation.}
\label{outline-analysis-table-TGCI} 
\small 
\renewcommand{\arraystretch}{1}
\begin{tabular}[t]{||l|c||}
\toprule
& {Motor Execution / Motor Imagery}\\
\midrule
\makecell{Gesture \\ to Perform}& 
\begin{minipage}{4.5in} 
\raggedright
Let $X$ and $Y$ represent two nodes in the EEG network. Compare whether $X$ is a Threshold Granger Cause $Y$ using the proposed approach given in Section \ref{proposed}.
\begin{enumerate}
\setlength\itemsep{0em}
\item Follow the steps as outlined in the TCI calculation procedure.
\item For each delay value \( d \) where threshold nonlinearity is detected, apply the Wald test to assess the predictive power of the threshold variable \( X_{t-d} \).
\item Combine the Wald test results over approximately 23 epochs for each subject.
\item Calculate the TGCI by averaging the Wald test results across all subjects and normalizing by the number of delays and subjects.
\item Visualize the TGCI results to identify the direction and strength of causal influence between EEG channels.
\end{enumerate}
\end{minipage}  \\
\bottomrule
\end{tabular}
\end{table*}

\subsection{Connectivity at Motor Imagery}

\begin{figure*}[t]
  \centering
  \subfloat[TCI at MI - Both Fists (BHs)\label{fig:tci_bh_mi}]{
    \includegraphics[width=0.45\textwidth]{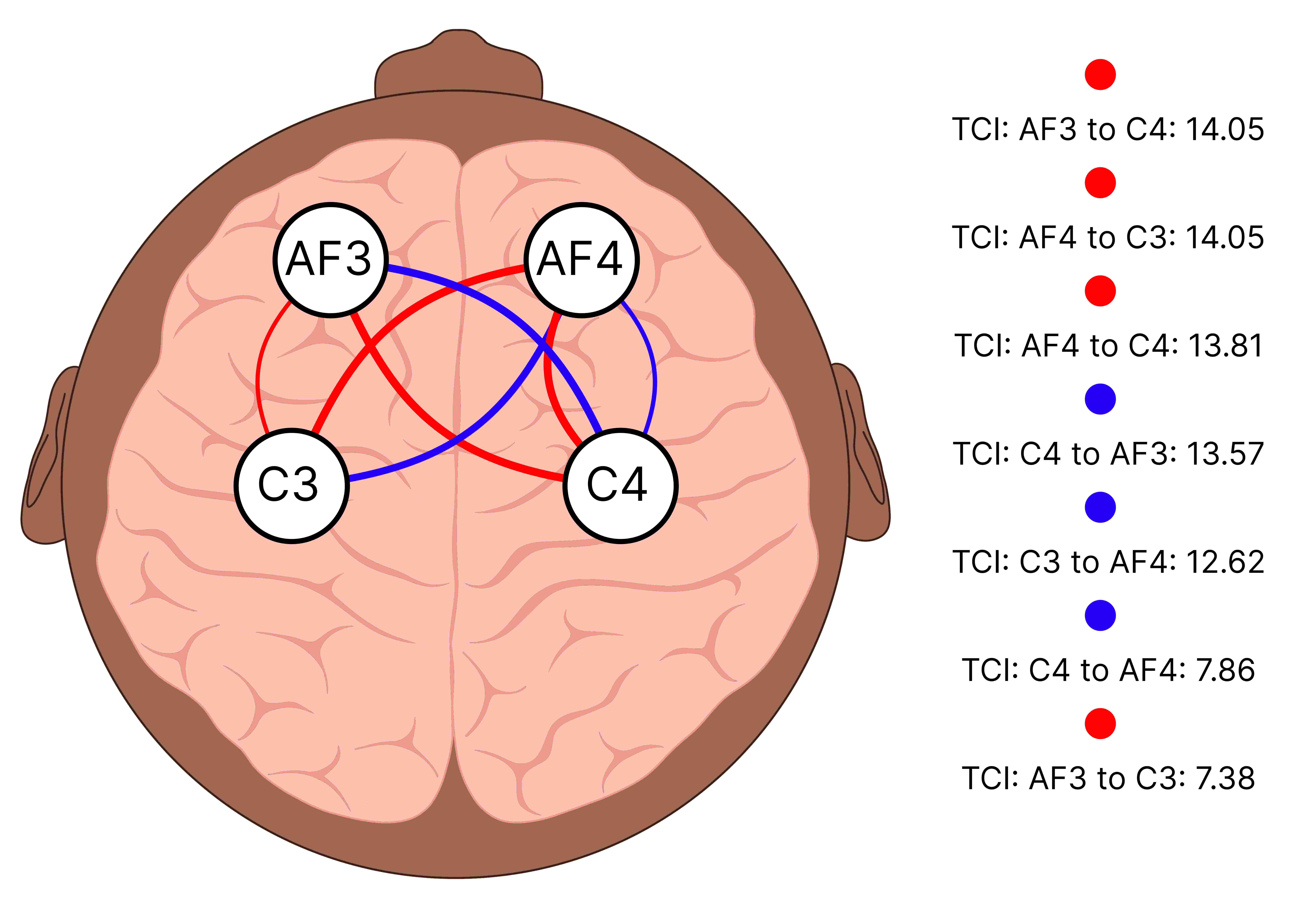}
  }\hfill
  \subfloat[TCI at MI - Both Feet (BF)\label{fig:tci_bf_mi}]{
    \includegraphics[width=0.45\textwidth]{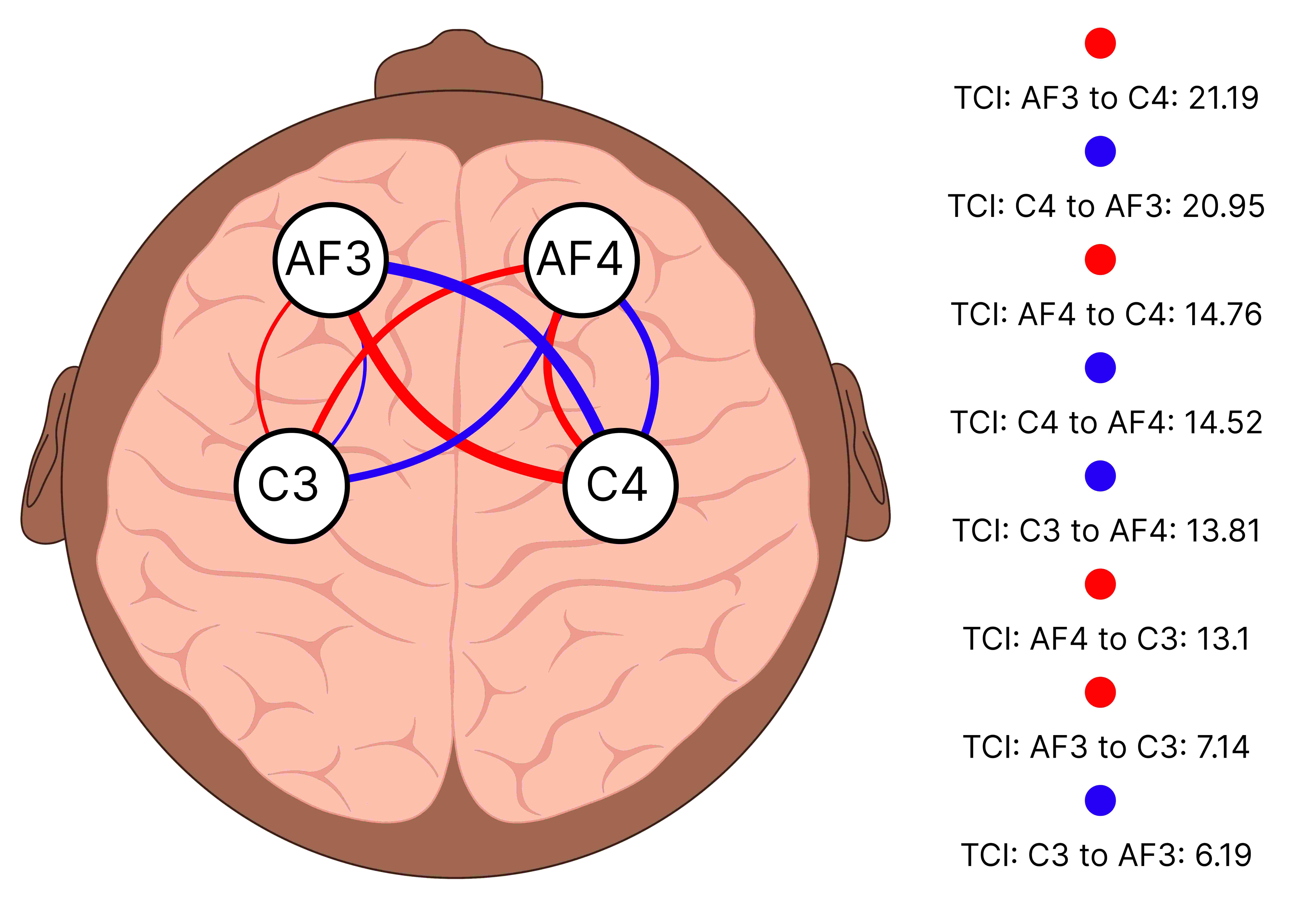}
  }\\[0.6em]
  \subfloat[TGCI at MI - Both Fists (BHs)\label{fig:tgci_bh_mi}]{
    \includegraphics[width=0.45\textwidth]{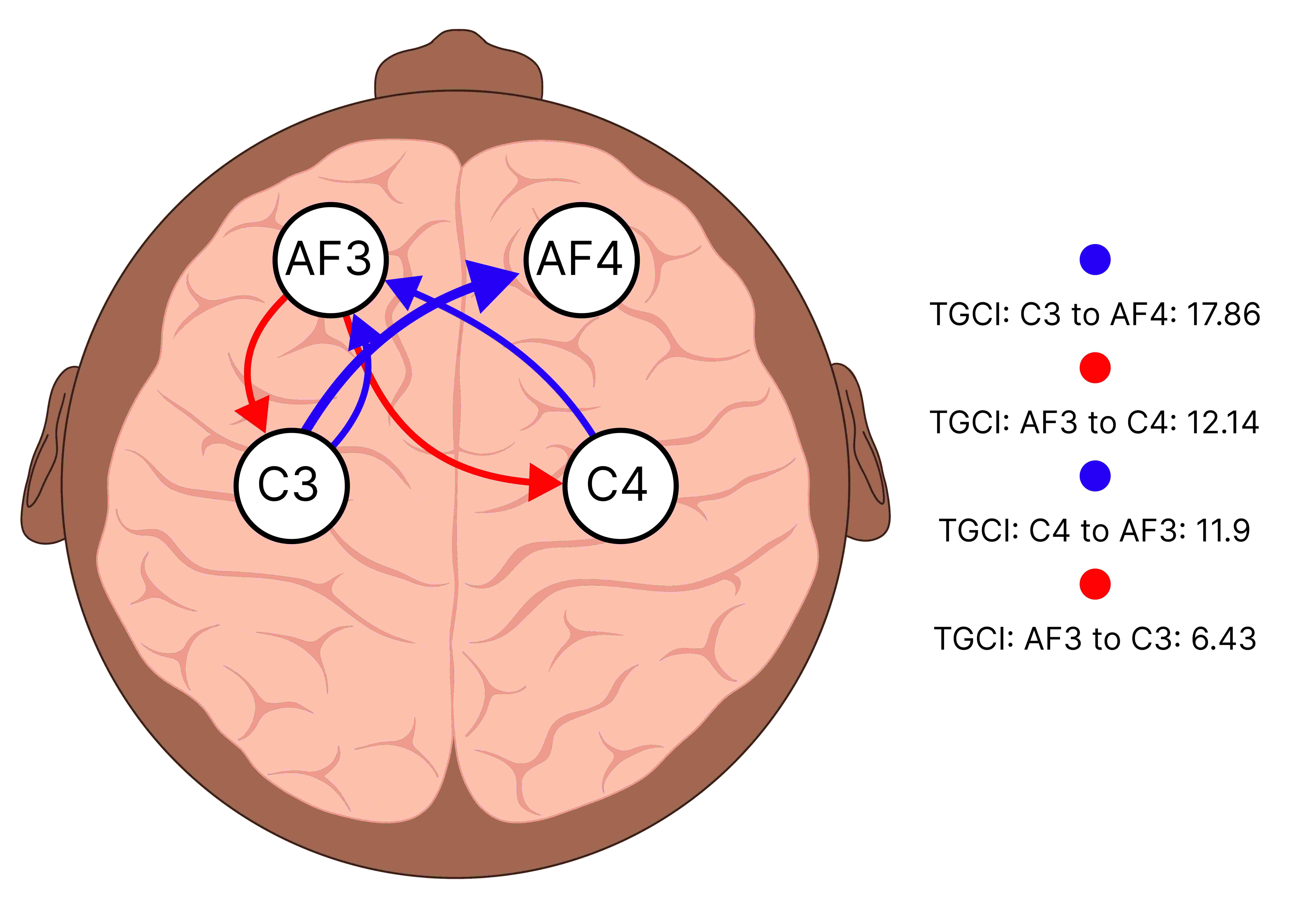}
  }\hfill
  \subfloat[TGCI at MI - Both Feet (BF)\label{fig:tgci_bf_mi}]{
    \includegraphics[width=0.45\textwidth]{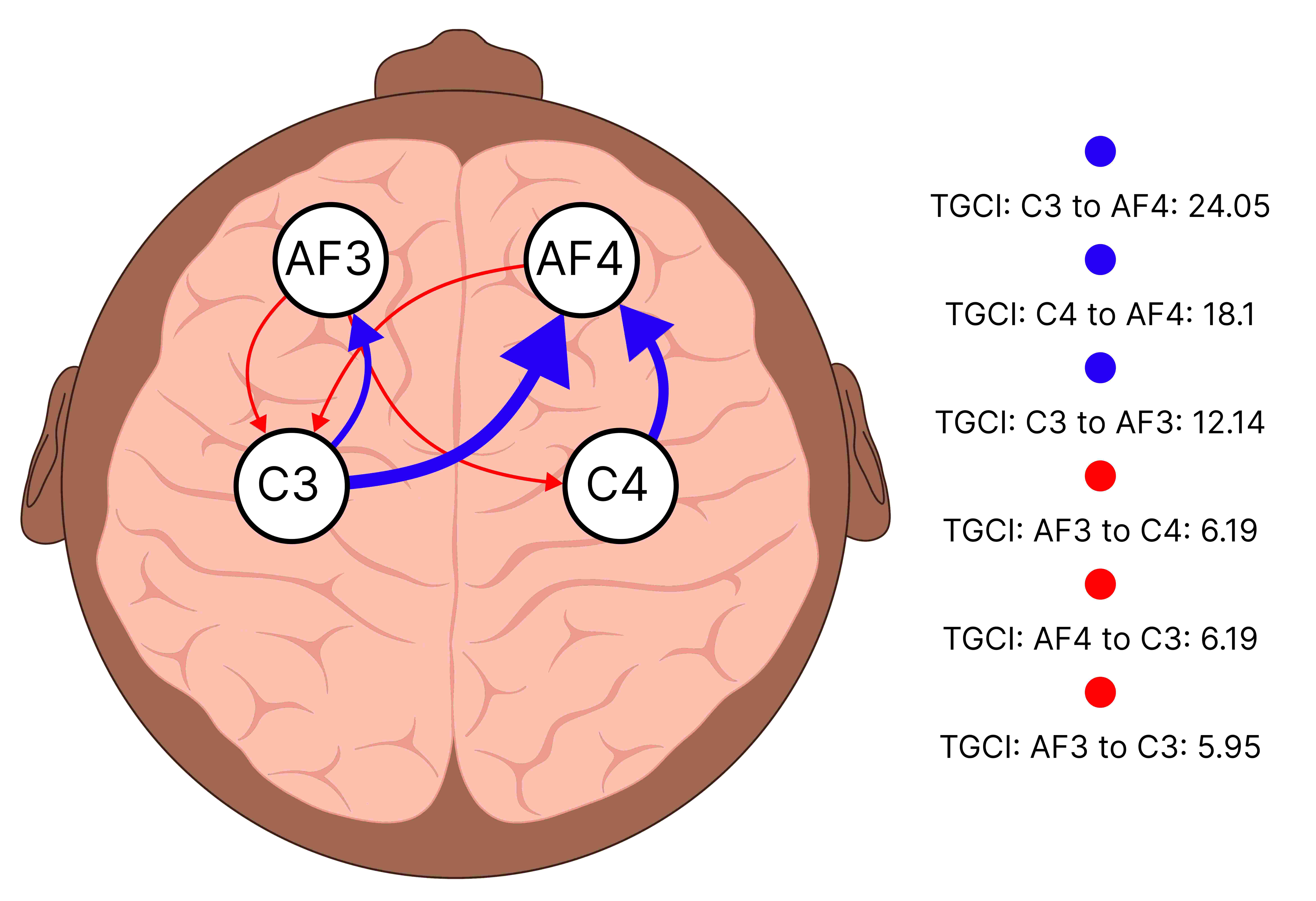}
  }

  \caption{Connectivity Nets at Motor Imagery. Panels: (a) TCI—BHs, (b) TCI—BF, (c) TGCI—BHs, (d) TGCI—BF.}
  \label{MI_TCI_TGCI}
\end{figure*}

\underline{\textbf{MI both hands' fists gesture (MI-BHs):}} 
For the MI-BHs gesture, the TCI results (Figure \ref{MI_TCI_TGCI}) show balanced and strong connectivity between the channels. The TCI value for AF3 $\to$ C3 is 7.38, while C3 $\to$ AF3 is 6.90. Meanwhile, AF3 $\to$ C4 is 14.05; C4 $\to$ AF3 is 13.57, indicating nearly equal connectivity in both directions. The connectivity involving AF4 is similarly strong and balanced, with TCI value from AF4 $\to$ C3 to be 14.05 and C3 $\to$ AF4 to be 12.62; AF4 $\to$ C4 is 13.81, C4 $\to$ AF4 is 7.86. The TGCI values for MI-BHs reveal significant contralateral causal influence in particular channels: C3 $\to$ AF4 17.86, AF3 $\to$ C4 12.14. The connection from C3 $\to$ AF4 and having no flow in the opposite direction of itself indicate a strong causal influence from the channel over the left motor to the right prefrontal cortex.

\underline{\textbf{MI both feet gesture (MI-BF):}} 
The MI-BF gesture (Figure \ref{MI_TCI_TGCI}) shows the highest TCI values, particularly for AF3 $\leftrightarrow$ C4:  AF3 $\to$ C4 21.19, C4 $\to$ AF3, 20.95, indicating strong and nearly equal connectivity in both directions. The connection AF4 $\leftrightarrow$ C3 and C4 also show high TCI values: AF4 $\to$ C3 13.10 and C3 $\to$ AF4 13.81; AF4 $\to$ C4 14.76, C4 $\to$ AF4 14.52, indicating strong and balanced connectivity. The TGCI results further emphasize this strong connectivity, with particularly high values for the C3 $\to$ AF4 24.05 and  C4 $\to$ AF4 18.10. These findings indicate that during the imagery of both feet task, communication between the channels continues, with the motor regions having a significant causal influence over the prefrontal areas. This prominent bidirectional interconnectivity appears to be an important element for the appropriate coordination of bilateral feet movement.

\underline{\textbf{Comparative analysis between gestures:}} 
The MI-BF gesture demonstrates stronger connectivity than MI-BHs, particularly between AF3 and C4. TCI values in MI-BF are higher (21.19 and 20.95 vs. 14.05 and 13.57 in BHs), reflecting more intense communication during MI-BF, with an increased role for the motor cortex. MI-BF highlights motor-to-prefrontal feedback, especially from [C3, C4] $\to$ AF4, pointing to stronger motor regulation of prefrontal activity during MI-BF. MI-BF also reveals more pronounced cross-hemispheric feedback between channels, suggesting denser bilateral engagement. Both tasks display strong bidirectional connectivity, with AF3 and AF4 affecting motor channels and vice versa. MI-BF suggests stronger motor-to-prefrontal feedback and overall higher connectivity, implying a more prominent dependence on motor cortex regulation during MI-BF.

\subsection{Connectivity at Motor Execution}
\begin{figure*}
  \centering
  \subfloat[TCI at ME - Both Fists (BHs)\label{fig:tci_bh_me}]{
    \includegraphics[width=0.45\textwidth]{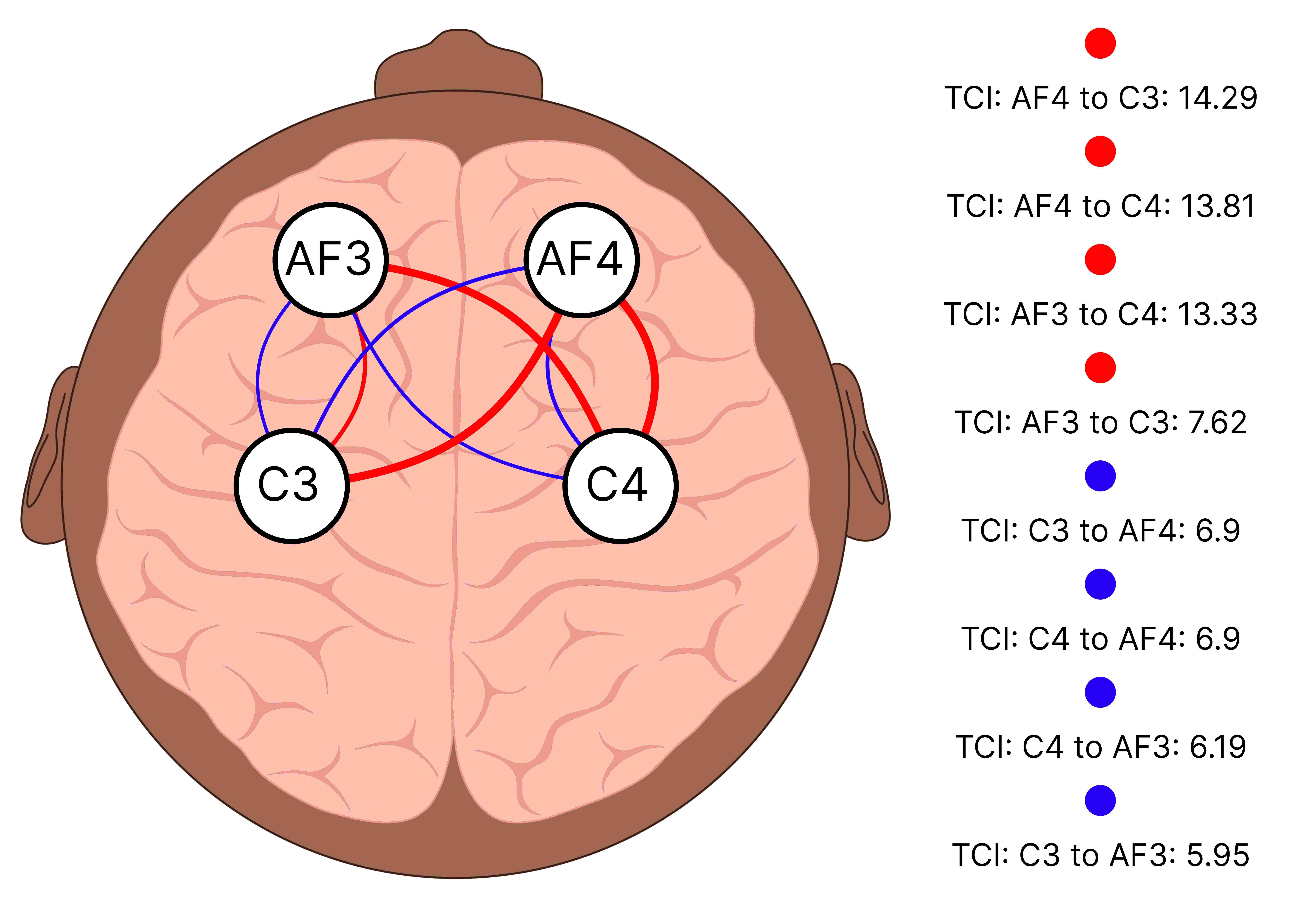}
  }\hfill
  \subfloat[TCI at ME - Both Feet (BF)\label{fig:tci_bf_me}]{
    \includegraphics[width=0.45\textwidth]{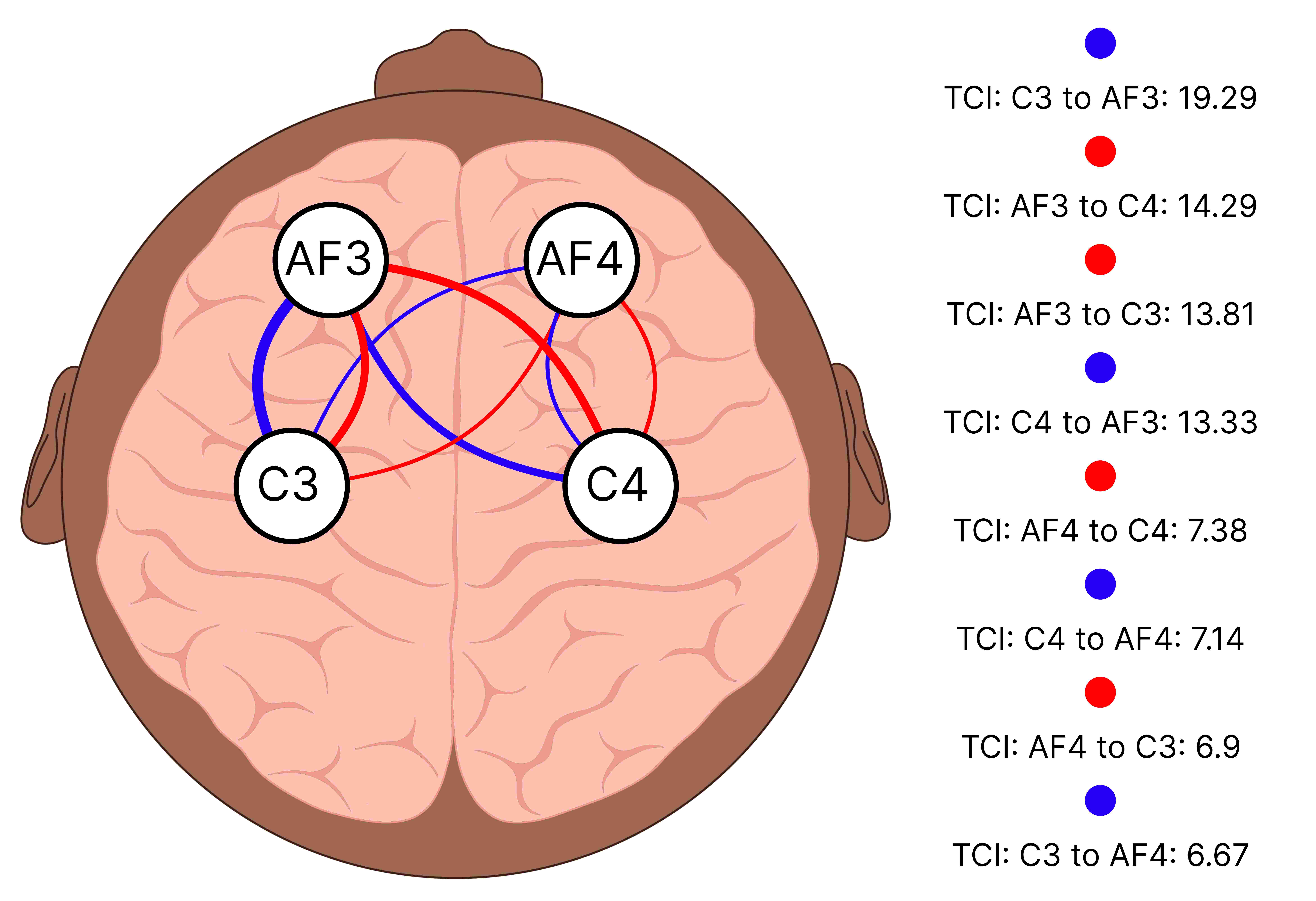}
  }\\[0.6em]
  \subfloat[TGCI at ME - Both Fists (BHs)\label{fig:tgci_bh_me}]{
    \includegraphics[width=0.45\textwidth]{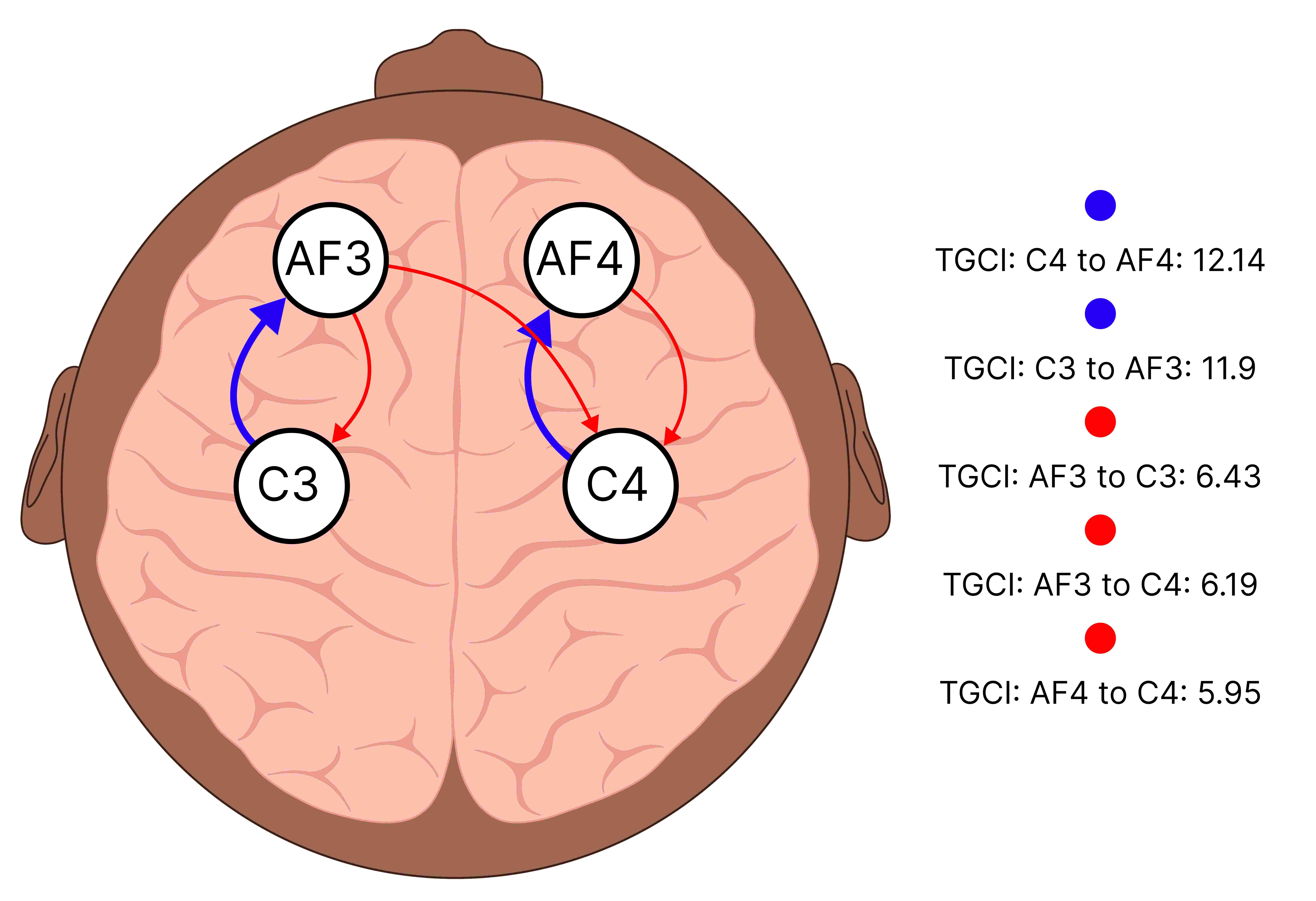}
  }\hfill
  \subfloat[TGCI at ME - Both Feet (BF)\label{fig:tgci_bf_me}]{
    \includegraphics[width=0.45\textwidth]{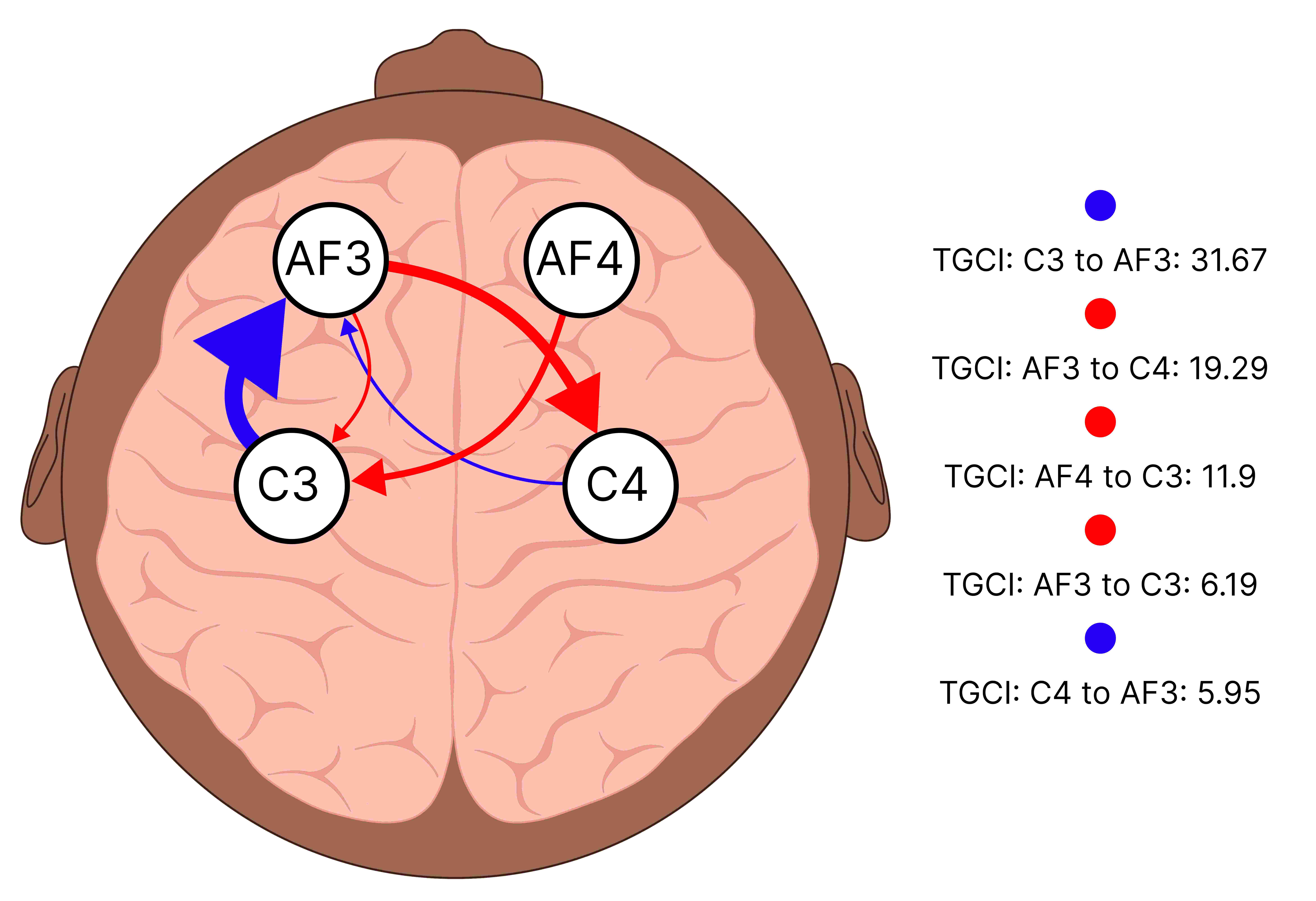}
  }

  \caption{Connectivity Nets at Motor Execution. Panels: (a) TCI—BHs, (b) TCI—BF, (c) TGCI—BHs, (d) TGCI—BF.}
  \label{ME_TCI_TGCI}
\end{figure*}

\underline{\textbf{ME both hand's fists gesture (ME-BHs):}} 
For MI-BHs gesture, the TCI results (Figure \ref{ME_TCI_TGCI}) show relatively proportional connectivity across the prefrontal and motor cortices. The TCI values from AF4 $\to$ C3 is 14.29, and from AF3 $\to$ C4, it is 13.33, suggesting noteworthy connectivity over C3 and C4. However, the TGCI results for MI-BHs gesture seem to oppose this finding, with significant causal influence observed from C4 $\to$ AF4 12.14 and from C3 $\to$ AF3 11.90. The contrast in terms of TCI and TGCI during the MI-BHs task suggests there is a continuous exchange of information between the channels, which might be related to coordinating bilateral motor activity. The bidirectional causal connectivity indicates that both hemispheres are actively engaged in the process.

\underline{\textbf{Motor execution -  both feet gesture (ME-BF):}} 
The ME both feet gesture shows the highest TCI values (Figure \ref{ME_TCI_TGCI}) among all gestures, particularly for C3 $\to$ AF3 19.29, and AF3 $\to$ C4 14.29. These results suggest that ME-BF involves strong connectivity between the channels. The TGCI results for ME-BF support this interpretation, showing very high TGCI values, especially C3 $\to$ AF3 31.67. This indicates a principal causal influence of the C3 over the AF3 during ME-BF, highlighting the significant feedback mechanisms necessary for executing and adjusting bilateral feet execution. A strong influence was observed AF3 $\to$ C4 19.29, reflecting the prefrontal cortex's role in the execution of this bilateral movement.

\underline{\textbf{Comparative analysis between gestures:}}
The ME-BF gesture shows stronger connectivity than ME-BHs, particularly in motor-to-prefrontal feedback. Higher TCI values (C3 $\to$ AF3: 19.29 vs. 5.95; C4 $\to$ AF3: 13.33 vs. 6.19) indicate greater motor cortex control over prefrontal areas during ME-BF, especially in the left hemisphere. In contrast, ME-BHs show weaker motor-to-prefrontal feedback, with more prefrontal control over the motor cortex. Both ME-BHs and ME-BF gestures involve prefrontal control over motor activity, but motor-to-prefrontal feedback is stronger in ME-BF, particularly from C3 $\to$ AF3. Cross-hemispheric connectivity is significant in both gestures, but stronger in ME-BF, particularly between C3, AF4, C4, and AF3. This reflects the greater bilateral coordination needed for ME-BF compared to ME-BHs movements. Both gestures show bidirectional connectivity between the prefrontal and motor cortex, with stronger motor-to-prefrontal feedback in ME-BF. The important difference is the dominant role of the motor cortex in ME-BF, especially in motor-to-prefrontal regulation, and stronger cross-hemispheric interactions during ME-BF.

\subsection{Comparative analysis of threshold Granger causality: imagery vs. execution}
\begin{figure*}[t]
  \centering
  \subfloat[Differences at MI - BHs\label{fig:tgc_bh_mi}]{
    \includegraphics[trim={13cm 69cm 18cm 27cm},clip,width=0.30\textwidth]{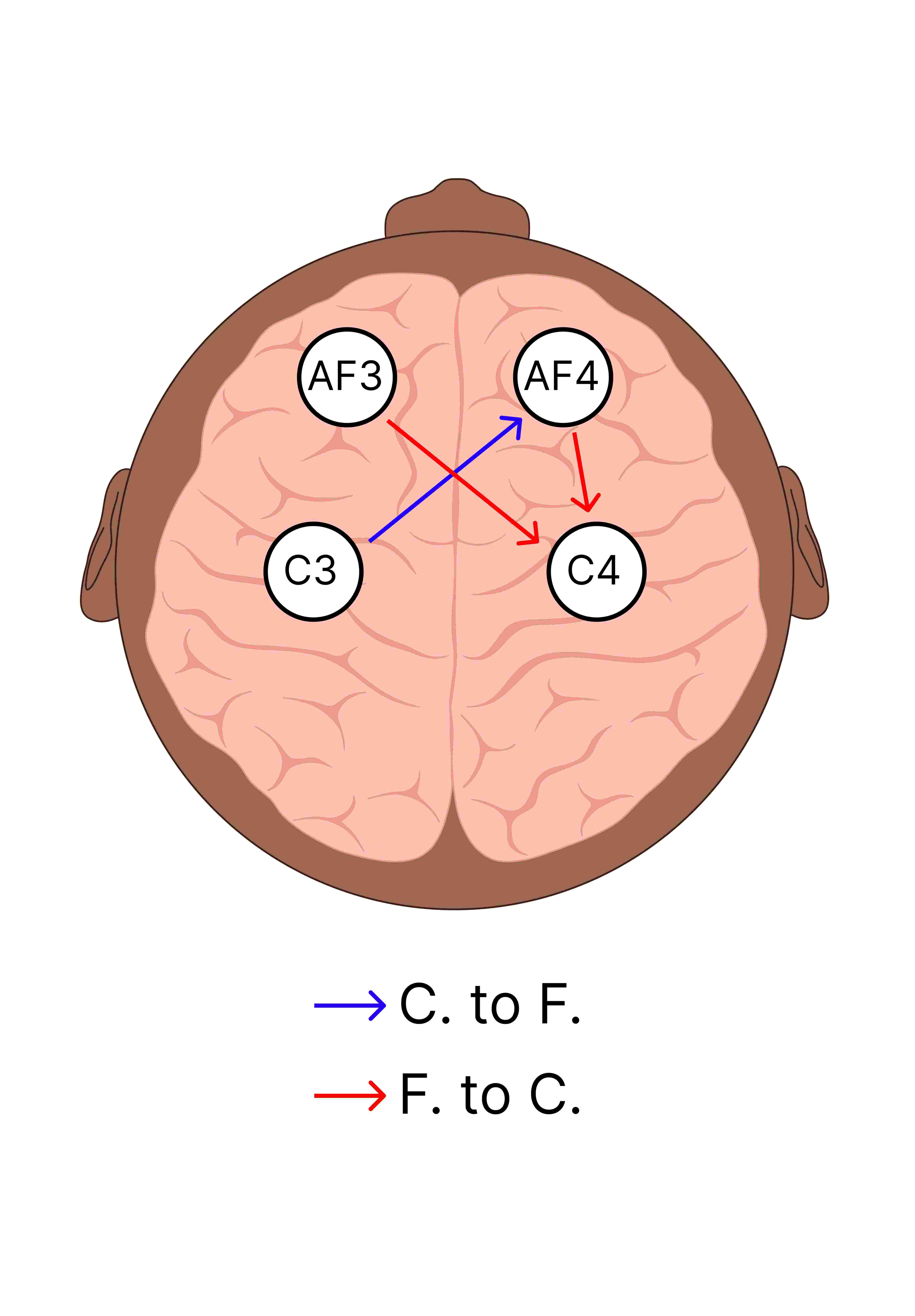}
  }\hfill
  \subfloat[Differences at ME - BHs\label{fig:tgc_bh_me}]{
    \includegraphics[trim={13cm 69cm 18cm 27cm},clip,width=0.30\textwidth]{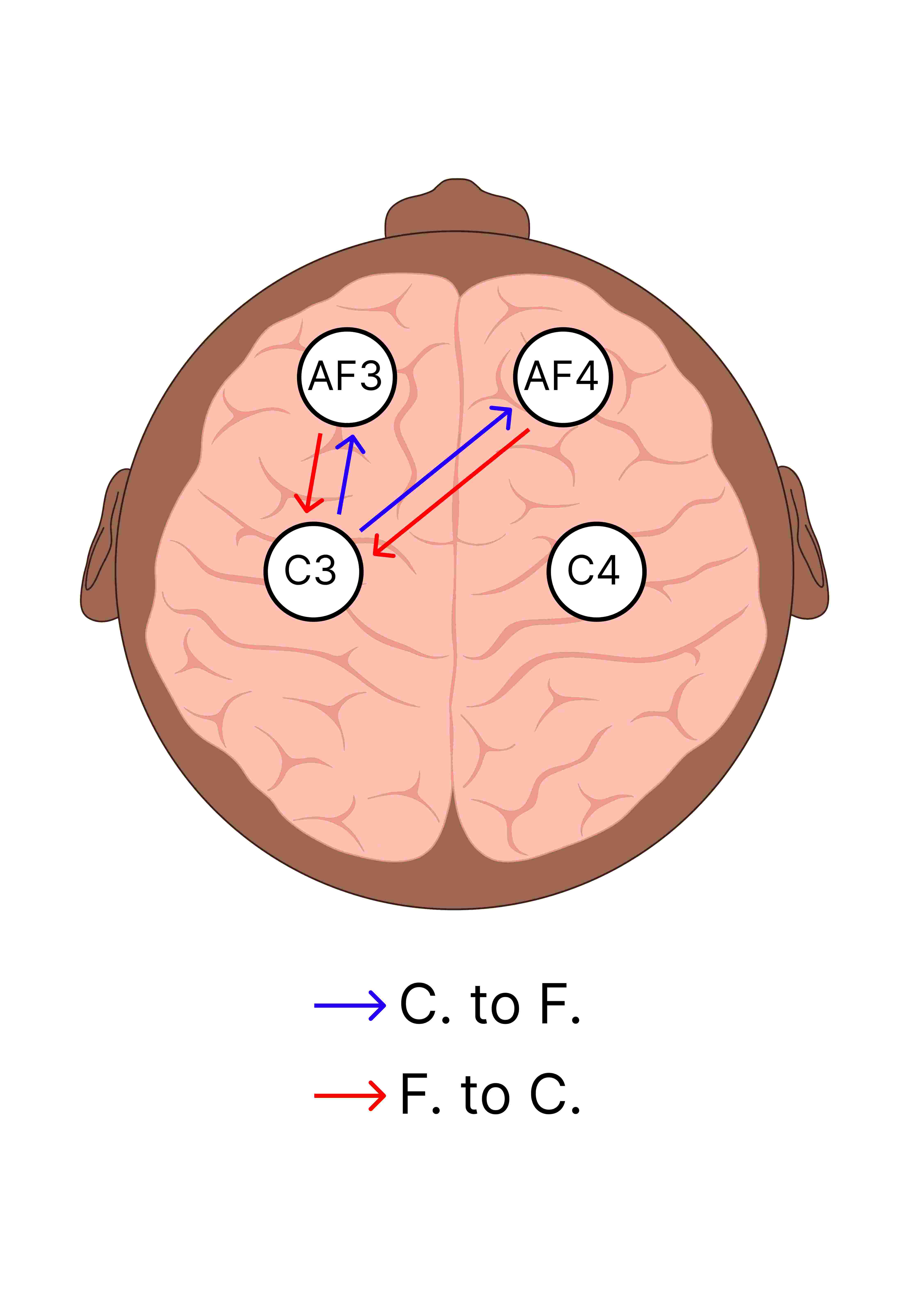}
  }\hfill
  \subfloat[Differences MI vs ME - BHs\label{fig:tgc_bh_mi_vs_me}]{
    \includegraphics[trim={13cm 69cm 18cm 27cm},clip,width=0.30\textwidth]{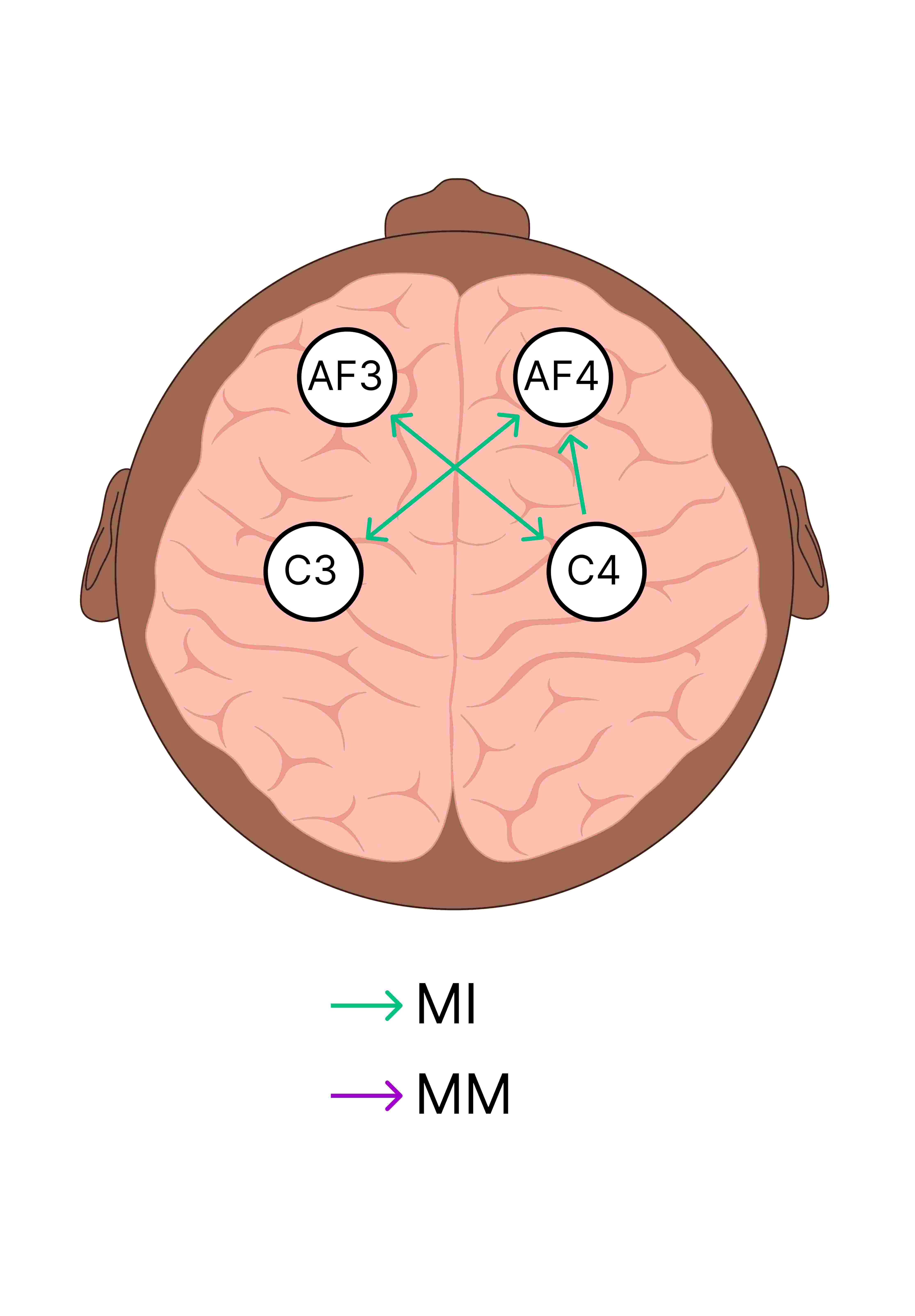}
  }\\[0.6em]
  \subfloat[Differences at MI - BF\label{fig:tgc_bf_mi}]{
    \includegraphics[trim={13cm 69cm 18cm 27cm},clip,width=0.30\textwidth]{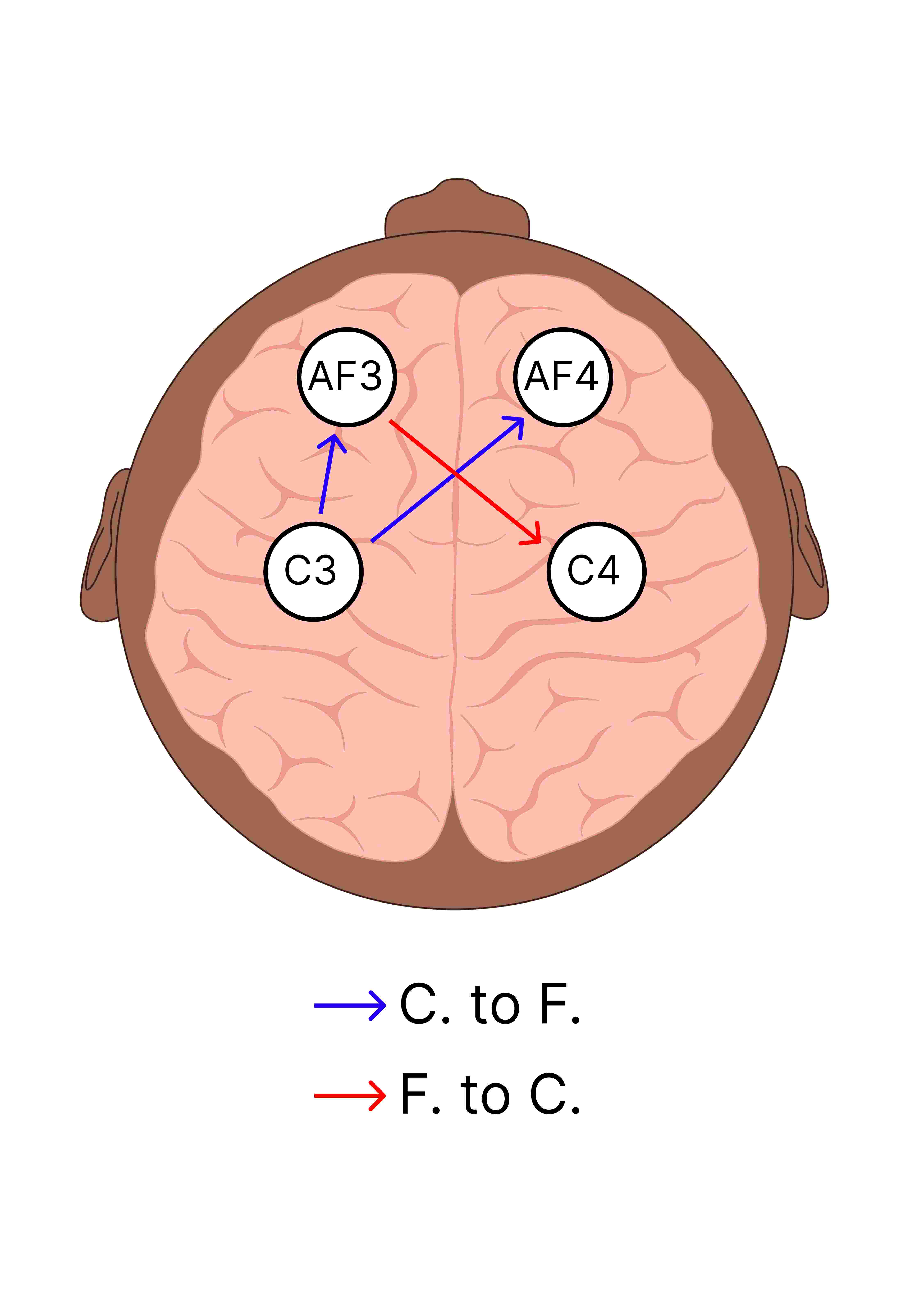}
  }\hfill
  \subfloat[Differences at ME - BF\label{fig:tgc_bf_me}]{
    \includegraphics[trim={13cm 69cm 18cm 27cm},clip,width=0.30\textwidth]{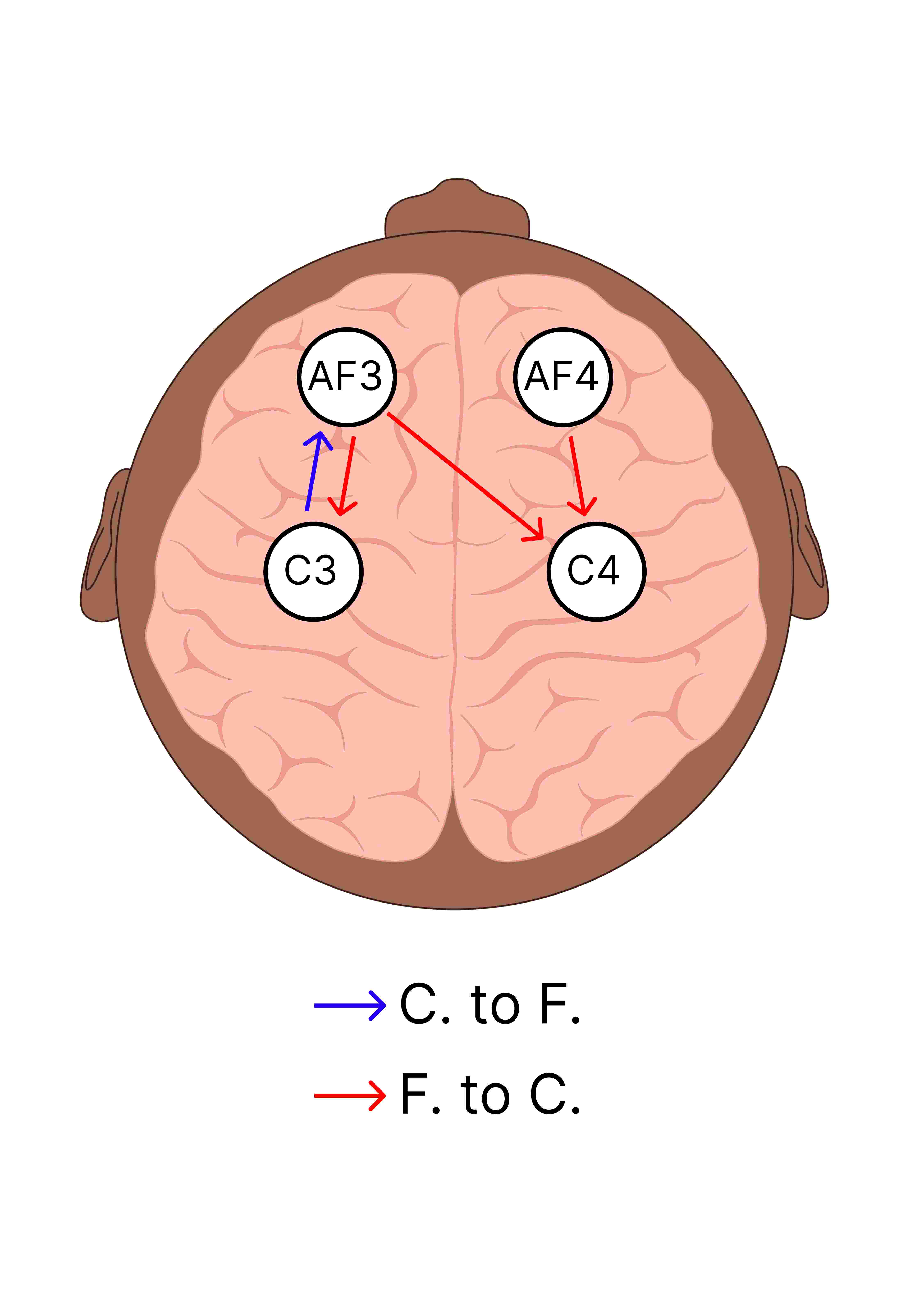}
  }\hfill
  \subfloat[Differences MI vs ME - BF\label{fig:tgc_bf_mi_vs_me}]{
    \includegraphics[trim={13cm 69cm 18cm 27cm},clip,width=0.30\textwidth]{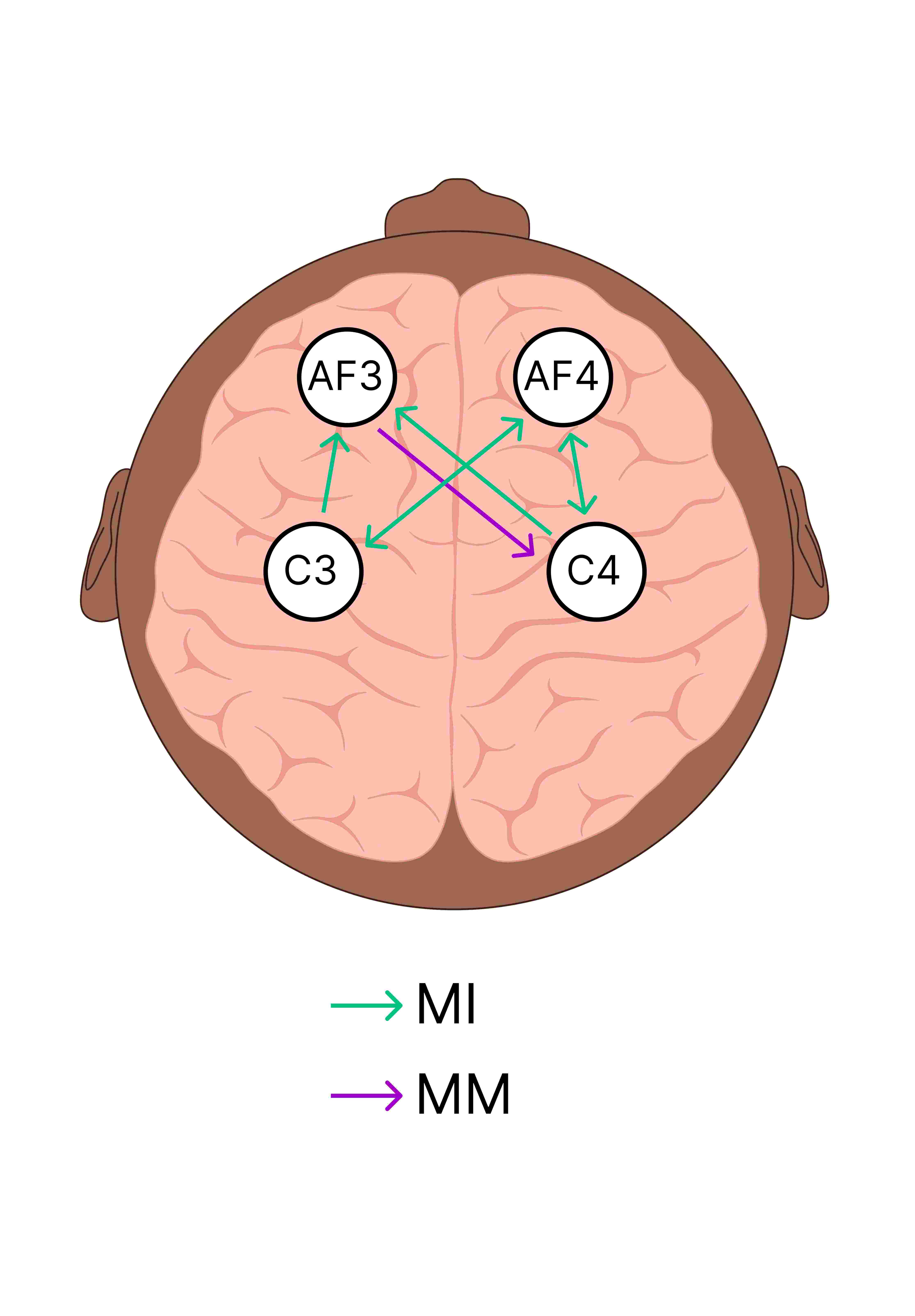}
  }

  \caption{Threshold Granger Causality (TGC) differences for Motor Imagery (MI), Motor Execution (ME), and MI vs.\ ME. Panels (a–c) show BHs; panels (d–f) show BF.}
  \label{fig:Comparisons_at_TGC}
\end{figure*}

We have thus far discussed the connectivity for MI and ME using TCI and TGCI values, as visualized in Figures \ref{MI_TCI_TGCI}  and \ref{ME_TCI_TGCI}. Further, we performed a permutational Hotelling's $T^2$ test (\cite{Hotelling}) on data from 35 subjects to examine causal connectivity differences within and between MI and ME tasks. For each pair, connectivity in one direction (e.g., $X$ $\to$ $Y$) was compared to the reverse direction ($Y$ $\to$ $X$) within MI and ME tasks. Significant differences are shown in Figure \ref{fig:Comparisons_at_TGC}, with columns 1 and 2 highlighting MI and ME differences, and column 3 showing differences between MI and ME in the same direction ($X$ $\to$ $Y$). Each arrow represents a statistically significant directional difference: red for prefrontal-to-motor, blue for motor-to-prefrontal, green for MI prominence, and purple for ME prominence. The comparison procedure is outlined in Table \ref{outline-analysis-table-TGC}:

\begin{table*}[ht]
\centering
\caption{Outline of the Implementation for Comparative Analysis of TGC differences}
\label{outline-analysis-table-TGC} 
\small
\renewcommand{\arraystretch}{1} 
\begin{tabular}[t]{||l|c||}
\midrule
\makecell{ Gesture \\ to Perform}& 
\linespread{0.75}\selectfont
\begin{minipage}[c]{4.5in}
      \raggedright
      For all subjects; let $X$ and $Y$ represent two nodes in the EEG network.
      \begin{enumerate}
        \item Apply the threshold nonlinearity test using the TAR(12) model for each epoch and combine test results. Then, generate a 35 (i.e., number of subjects) by 1 vector for each delay \(d\).
          \begin{enumerate}
            \item Retain the F-test statistics for subjects where linearity is rejected and assign 0s for others.
          \end{enumerate}
        \item Apply the Wald test on the data of subjects who reject linearity to assess the predictive power of the threshold variable \(X_{t-d}\).
          \begin{enumerate}
            \item Filter the vector of F-test statistics based on subjects who pass the Wald test.
            \item Retain only those elements corresponding to subjects who pass the Wald test, forming the TGC distribution for each delay \(d\).
          \end{enumerate}
        \item Compare the TGC distributions for each pairwise channel interaction \(X \to Y\) vs. \(Y \to X\) within the same task (MI or ME) using the permutational Hotelling’s \(T^2\) test.
          \begin{enumerate}
            \item Identify statistically significant differences in the directional connectivity within the same task.
            \item Visualize significant differences with arrows:
              \begin{itemize}
                \item \textbf{Red} = prefrontal \(\to\) motor.
                \item \textbf{Blue} = motor \(\to\) prefrontal.
              \end{itemize}
          \end{enumerate}
        \item Compare TGC in the same direction (e.g., \(X \to Y\)) across MI vs. ME with the permutational Hotelling’s \(T^2\) test.
          \begin{enumerate}
            \item Flag significant MI–ME differences per pair.
            \item Visualize significant differences:
              \begin{itemize}
                \item \textbf{Green} = stronger at MI.
                \item \textbf{Purple} = stronger at ME.
              \end{itemize}
          \end{enumerate}
      \end{enumerate}
    \end{minipage}\\
\bottomrule
\end{tabular}
\end{table*}

For the MI-BHs, Figure \ref{fig:Comparisons_at_TGC} highlights significant connectivity from AF3 $\to$ C4 and AF4 $\to$ C4, which is consistent with the TGCI results showing similar connections 12.14 for AF3 $\to$ C4. This alignment between the significant differences in TGC samples and TGCI underlines the importance of AF3 and AF4 in coordinating BHs movement during MI. Figure \ref{fig:Comparisons_at_TGC} also identifies significant motor-to-prefrontal feedback for MI-BHs from C3 $\to$ AF4, and this is confirmed by TGCI results, which show a strong connection from C3 $\to$ AF4 17.86. The agreement between the two methods suggests robust bidirectional dynamics during MI-BHs, with information flow amongst channels.

In the ME-BHs, Figure \ref{fig:Comparisons_at_TGC} identifies connectivity from AF3 $\to$ C3 and AF4 $\to$ C3. This finding is consistent with the TGCI results, which also highlight connections from AF3 $\to$ C4 6.19 and AF4 $\to$ C4 5.95. These connections emphasize the critical role of AF3 and AF4 in coordinating the ME-BHs. However, the blue arrows from C3 $\to$ AF3 and C3 $\to$ AF4 in Figure \ref{fig:Comparisons_at_TGC} indicate significant feedback from C3 $\to$ AF3 and $\to$ AF4, which are partially captured by TGCI results, such as the connection from C4 $\to$ AF3 11.9. This feedback highlights the bidirectional nature of the connectivity necessary for maintaining coordination during bilateral fists task. The combination of TGC sample differences and TGCI results highlights the condition for continuous, reciprocal communication between the prefrontal and motor regions to achieve coordinated motor execution.

For the MI-BF, Figure \ref{fig:Comparisons_at_TGC} shows significant differences from C3 $\to$ AF3 and C3 $\to$ AF4. These connections indicate that C3 influences both the left and right prefrontal area channels during this task. Additionally, Figure \ref{fig:Comparisons_at_TGC} reveals a significant connection between AF3 and C4, indicating that the AF3 influence over C4 was evident during bilateral MI-BF. Comparing these findings with the TGCI results, we observe a strong connection from C3 $\to$ AF3 12.14 and from C3 $\to$ AF4 24.05, reinforcing the importance of these pathways during MI. There is also a consistent outcome with regard to the connectivity between AF3 and C4 6.19(as indicated by the red arrow in Figure \ref{fig:Comparisons_at_TGC}). The combination of both emphasizes the bidirectional dynamics necessary for coordinating complex motor imagery tasks involving both feet.

In contrast, for the ME-BF, Figure \ref{fig:Comparisons_at_TGC} shows significant connectivity from AF3 $\to$ C3, AF3 $\to$ C4, and AF4 $\to$ C4, highlighting the prominent role of AF3 and AF4 in driving motor activity during bilateral ME-BF. Additionally, significant feedback is observed from C3 $\to$ AF3. TGCI results for the ME-BF further emphasize this feedback, particularly highlighting the strong connection from C3 $\to$ AF3 31.67, which is the most pronounced feedback pathway observed. The consistency between the significant differences in TGC samples and TGCI findings suggests that both methods effectively capture the critical bidirectional flow of information between the channels over the prefrontal and motor areas during ME. The significant connections identified by TGCI (e.g., AF3 $\to$ C4 with TGCI = 19.29, and AF4 $\to$ C3 with TGCI = 11.9) further support the argument that AF3 and AF4 take part in the bilateral ME-BF.

\underline{\textbf{Comparison: imagery vs. execution}} 
Figure \ref{fig:Comparisons_at_TGC}, 1st-row 3rd-column shows a clear preference in favor of MI-BHs regarding bidirectional connectivity, as indicated by the green arrows from AF3 $\to$ C4, AF4 $\to$ C3, C3 $\to$ AF4, C4 $\to$ AF3, and C4 $\to$ AF4. These findings suggest that MI requires a more integrated and reciprocal interaction between channels to coordinate bilateral fists movements. The TGCI results support this observation, with significant connectivity noted from C3 $\to$ AF4 17.86 and from C4 $\to$ AF3 11.90 during MI. Conversely, the absence of purple arrows in Figure \ref{fig:Comparisons_at_TGC} for both fists movement indicates that ME does not engage these bidirectional connections as strongly as MI, suggesting a more straightforward connectivity pattern during actual movement.

Figure \ref{fig:Comparisons_at_TGC}, 2nd-row 3rd-column highlights a strong preference for MI-BF, as evidenced by the green arrows indicating stronger bidirectional connectivity from AF4 $\to$ C3, AF4 $\to$ C4, C3 $\to$ AF3, C3 $\to$ AF4, C4 $\to$ AF3, and C4 $\to$ AF4. These results suggest that MI of BF involves complex interactions between the channels, likely due to the cognitive demands of coordinating bilateral feet movement. The TGCI results corroborate this, with strong feedback from C3 $\to$ AF4 24.05 and from C4 $\to$ AF4 18.10 during MI. The purple arrow from AF3 $\to$ C4 in Figure \ref{fig:Comparisons_at_TGC}, 2nd-row 3rd-column suggests that ME involves stronger prefrontal-to-motor connectivity in this specific pathway, aligning with the TGCI result showing significant connectivity from AF3 $\to$ C4 19.29 during ME. This difference in connectivity patterns between MI and ME for BF motion tasks highlights the distinct neural circuitries employed for imagined versus executed, with MI requiring more extensive feedback across hemispheres.

Across tasks, MI shows richer and more two‑way connectivity than ME. Significant differences in the TGC samples and TGCI patterns point to distinct mechanisms for imagery and execution. The TGC samples are sensitive to bidirectional feedback and expose fine interactions that TGCI may weigh differently. TGCI, in turn, summarizes the net strength and direction of links and highlights the connections that support motor control and imagery. Taken together, these two views are complementary; any seemingly unrelated discrepancy reflects emphasis, not contradiction.

The results are consistent with prior work on MI and ME \citep{chen2009evaluation, vigasina2023eeg, drapkina2024characteristics}. Many studies have mapped EEG connectivity within sensorimotor and frontal systems. Using TAR models adds the ability to capture time-varying, nonlinear causal relations between channels. For example, \cite{solodkin2004fine} reported lower network complexity during MI than during ME; our analysis refines this view by tracking how connectivity changes over time and by delay. By testing state-aware delay‑specific effects and applying formal tests for causality, we provide a novel interpretation of dynamic connectivity. Methodologically, integrating temporal causality into network analysis improves current practice and provides useful insight into motor‑related neural dynamics.

\section{Conclusion}
\label{sec:conclusion}
This study devises a novel connectivity conceptualization called threshold connectivity aimed at exploring nonlinear and time-varying causal interactions in time series networks. TAR4C delivers an interpretable pathway to nonlinear and state‑aware causal inference in time series networks. The procedure is two‑stage: detect threshold connectivity, then test threshold Granger causality within driver‑governed regimes. Confounding background activity in the network is mitigated through spectral dynamic principal components before pairwise modeling, which improves inference on the nodes of interest. The computational cost is the trade‑off, but the representation is straightforward and testable.

The EEG study aimed to illustrate the relevance to real-world applications. Motor imagery exhibits denser and more reciprocal interactions than motor execution. Cross‑hemispheric links are prominent under MI, whereas ME reveals streamlined feedback, especially motor‑to‑prefrontal during bilateral feet movements. These contrasts emerge coherently from TCI/TGCI and the distributional comparisons of TGC.

Limitations can be addressed in the following ways. First, the modeling is pairwise and local. Network‑wide dependencies can persist even after confounding background removal and may modulate the identified links. Second, inference at the sensor level is constrained by mixing, volume conduction, and referencing. These factors can bias directionality and inflate near‑zero‑lag structure. Third, grid search over hyperparameters such as thresholds, delays, and orders is computationally heavy and risks over‑parameterization; we mitigate this with heteroskedasticity‑robust tests and a parsimonious model‑selection criterion. Fourth, the threshold switch is discrete in its current form; it does not imply a smooth transition or a system‑level bifurcation. However, the scope of threshold connectivity can be enriched with different regime-switching mechanisms and model specifications. Results should be interpreted as effective connectivity between signals, not as a direct implication of synaptic transmission.

Several research directions can be considered for methodological extensions. (i) Move from pairwise to vector/multivariate TAR to model small subnetworks jointly and reduce omitted‑variable bias. (ii) Introduce sparsity or graph‑guided penalties for high‑dimensional settings to stabilize estimation and variable selection. (iii) Allow multiple threshold functionals and smooth/logistic transitions to capture graded gating; estimate time‑varying thresholds and regime persistence. (iv) Pool information across subjects/units/repetitions via hierarchical or empirical‑Bayes structures to strengthen group‑level inference. (v) Combine pairwise outputs into network‑level tests with multiplicity control. (vi) Accelerate computation using techniques such as branch and bound search, dynamic programming, or GPU‑enabled arranged autoregressions. The multivariate generalization is already feasible within the framework. 

Regarding the brain dynamics specifically, the regime‑switching mechanism in TAR4C echoes basic motifs of phase transitions in the metastable brain principle and the critical‑brain hypothesis. TAR4C can therefore serve as a numerical probe for these theories when paired with appropriate diagnostics. With the evaluative program, TAR4C can support systematic, data‑driven assessment of the metastability principle and criticality hypothesis.

Beyond neuroscience, TAR4C can be applied in fields dealing with complex time series data, such as economics, finance, and climate studies. For example, in economics or finance, where one needs to understand the dynamic interactions of various economic indicators or several stock prices, it can identify threshold nonlinearity-dependent causal connections that may remain hidden while using conventional methods. Similarly, it can be used in climate science to study the causal relationships between meteorological variables in different locations.

In summary, TAR4C isolates threshold‑governed influence and exposes state‑dependent directionality in an interpretable, testable form. The proposed approach complements conventional methods by incorporating regime-switching controls to causal inference and formal statistics. Its limits are clear and tractable; its extensions are within reach. 

\section*{Data Availability}
The data that support the findings of this study are openly available in PhysioNet, at https://physionet.org/content/eegmmidb/1.0.0/, reference number https://doi.org/10.13026/C28G6P. Raw data were generated at the BCI R$\&$D Program, Wadsworth Center, New York State Department of Health, Albany,NY. W.A. Sarnacki collected the data. Aditya Joshi compiled the dataset and prepared the documentation. D.J. McFarland and J.R. Wolpaw were responsible for experimental design and project oversight, respectively. This work was supported by grants from NIH/NIBIB ((EB006356 (GS) and EB00856 (JRW and GS)).
\section*{References}
\nocite{*}
\bibliography{Bibliography-TAR4C}

\end{document}